\documentclass{WileyMSP-template}
\graphicspath{{Figures}}
\usepackage{upgreek}
\usepackage{hyperref}
\usepackage[dvipsnames]{xcolor}
\usepackage{xr}
\begin{document}

\pagestyle{fancy}
\rhead{\includegraphics[width=2.5cm]{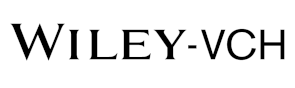}}

\title{A large scale multi-modal workflow for battery characterization: from concept to implementation}

\maketitle

% Author: Please give full first and last names for authors and include * after the name of all corresponding authors

\author{François Cadiou$^{\ 1,2,}$*,}
\author{Cinthya Herrera$^{3,4}$,}
\author{Duncan Atkins$^{3}$,}
\author{Elixabete Ayerbe$^{5}$,}
\author{Giorgio Baraldi$^{5,6}$,}
\author{Stéphanie Belin$^{7}$,}
\author{Anass Benayad$^{2}$,}
\author{Didier Blanchard$^{1}$,}
\author{Federico Capone$^{8,9}$,}
\author{Ennio Capria$^{1}$,}
\author{Isidora Cekic-Laskovic$^{10}$,}
\author{Robert Dominko$^{11}$,}
\author{Kristina Edström$^{12}$,}
\author{Ajay Gautam$^{13}$,}
\author{Lukas Helfen$^{3}$,}
\author{Antonella Iadecola$^{8,9}$,}
\author{Quentin Jacquet$^{14}$,}
\author{Gregor Kapun$^{11}$,}
% \author{Francesco La Porta$^{7}$,}
\author{Xinyu Li$^{15}$,}
\author{Aleksandar Matic$^{16}$,}
\author{Nataliia Mozhzhukhina$^{16}$,}
\author{Andrew J. Naylor$^{12}$,}
\author{Poul Norby$^{15}$,}
\author{Chris O’Keefe$^{17}$,}
% \author{Luc\'ia P\'erez Ram\'irez$^{7}$,}
\author{Alexandre Ponrouch$^{18}$,}
\author{Jean-Pascal Rueff$^{7}$,}
\author{Elena Tchernykova$^{11}$,}
\author{Deyana Tchitchekova$^{18}$,}
\author{Israel Temprano$^{17}$,}
\author{Nikita Vostrov$^{1}$,}
\author{Marnix Wagemaker$^{13}$,}
\author{Martin Winter$^{10}$,}
% \author{Gilles Wittmann$^{7}$,}
\author{Christian Wölke$^{10}$,}
\author{Tejs Vegge$^{15}$,}
\author{Sandrine Lyonnard$^{\ 14,}$*}

% Dedication

% \dedication{Optional dedication here. If no dedication is required, please leave blank}

% Affiliations: Please provide academic titles (Prof. or Dr.) for all authors where applicable, and include an institutional email address for all corresponding authors
\begin{affiliations}
%Dr. François Cadiou, Dr. Nikita Vostrov, Dr. Didier Blanchard, Dr. Ennio Tito Capria\\
$^{1}$ European Synchrotron Radiation Facility, 71, avenue des Martyrs, CS 40220, 38043 Grenoble Cedex 9, France\\
Email Address: francois.cadiou@cea.fr

% Dr. François Cadiou, Prof. Anass Benayad\\
$^{2}$ Univ. Grenoble Alpes, CEA, Liten, 17 Rue des Martyrs, 38054 Grenoble Cedex, France

% Dr. Cinthya Herrera, Lukas Helfen, Dr. Duncan Atkins\\
$^{3}$ Institut Laue-Langevin, 71 avenue des Martyrs, CS 20156, 38042 Grenoble Cedex 9, France

% Dr. Cinthya Herrera\\
$^{4}$ Univ. Grenoble Alpes, CEA, IRIG-MEM-L\_SIM, 17 Rue des Martyrs, 38054 Grenoble Cedex, France

% Prof. Elixabete Ayerbe, Dr. Giorgio Baraldi\\
$^{5}$ CIDETEC, Basque Research and Technology Alliance (BRTA), Po. Miramón 196, 20014 Donostia-San Sebastián, Spain

% Dr. Giorgio Baraldi\\
$^{6}$ Basquevolt, Alava Technology Park, Albert Einstein 35, 01510, Vitoria-Gasteiz, Spain

% Dr. Jean-Pascal Rueff, Dr. Stéphanie Belin\\
$^{7}$ Synchrotron SOLEIL, L’Orme des Merisiers, Départementale 128, Saint Aubin 91190, France

% Dr. Federico Capone, Dr. Antonella Iadecola\\
$^{8}$ Sorbonne Université, CNRS, Physicochimie des Électrolytes et Nanosystèmes Interfaciaux, Paris F-75005, France

% Dr. Federico Capone, Dr. Antonella Iadecolla\\
$^{9}$ Réseau sur le Stockage Electrochimique de L’Energie (RS2E), FR CNRS 3459, Amiens 80039, France

% Prof. Isidora Cekic-Lascovic, Dr. Christian Wolke, Prof. Martin Winter\\
$^{10}$ Forschungszentrum Jülich GmbH, Helmholtz-Institute Münster (IMD-4), Corrensstraße 48, 48149 Münster, Germany

% Gregor Kapun, Dr. Elena Tchernykova, Prof. Robert Dominko\\
$^{11}$ National institute of Chemistry, Department of Materials Chemistry, Hajdrihova 19, Ljubljana 1000, Slovenia

% Dr Andrew J. Naylor, Prof. Kristina Edström\\
$^{12}$ Uppsala University, Department of Chemistry – Ångström Laboratory, SE751 21 Uppsala, Sweden

% Dr. Ajay Gautam, Prof. Marnix Wagemaker\\
$^{13}$ University of Technology Mekelweg 15 - Department of Radiation Science and Technology Delft - Delft 2629JB, THE NETHERLANDS

% Dr. Quentin Jacquet, Prof. Sandrine Lyonnard\\
$^{14}$ Univ. Grenoble Alpes, CEA, CNRS, GrenobleINP, IRIG-Symmes, 17 Rue des Martyrs, 38054 Grenoble Cedex, France\\
Email Address: sandrine.lyonnard@cea.fr

% Dr. Xinyu Li, Prof. Poul Norby, Prof. Tejs Vegge\\
$^{15}$ Technical University of Denmark (DTU), Department of Energy Conversion and Storage, Anker Engelunds Vej, Kgs. Lyngby DK-2800, Denmark

% Dr. Nataliia Mozhzkhina, Aleksandar Matic\\
$^{16}$ Chalmers University of Technology, Department of Physics, SE 412 96, Gothenburg, Sweden

% Dr. Chris O'Keefe, Dr. Israel Temprano\\
$^{17}$ University of Cambridge, Department of Chemistry, Cambridge CB2 1EW, England

% Dr. Alexandre Ponrouch, Dr. Deyana Tchitchekova\\
$^{18}$ Institut de Ciència de Materials de Barcelona,  ICMAB-CSIC, Campus UAB, 08193 Bellaterra, Spain

\end{affiliations}

% Keywords: Please provide a minimum of three and a maximum of seven keywords, separated by commas

\keywords{Battery, Workflow, Experimental characterization, Multi-modal, Large scale, Correlation, Aging}

% Abstract should be written in the present tense and impersonal style (i.e., avoid we), and be at most 200 words long
\begin{abstract}

The development of material acceleration platforms in battery research requires integrating complementary techniques and correlating heterogeneous experimental datasets. Here, this challenge is tackled in a large-scale multimodal program involving fifteen laboratories and facilities across Europe. Coordinated multi-site experiments are performed on state-of-the-art graphite~/ LiNiO$_{2}$ Li-ion full cells to address two archetypal scientific questions: is the electrolyte composition impacting electrode properties, and how do electrode materials evolve when cells are cycled to their end-of-life? A fully standardized and centralized workflow is demonstrated, from sample production and delivery, to metadata and data handling, generating seventy-five concatenated datasets shared among all partners. Their integrated analysis shows that scientific conclusions depend critically on both the observable chosen to describe electrode properties, and the measurement technique employed. Individual experiments provide detailed information into specific aspects, such as crystal structures, redox activity, surface processes, morphology, \textit{etc.} – but can also function as binary diagnostic tool. Two-dimensional observable-technique patterns are introduced, in which each pixel encodes a yes, no or uncertain answer to a given scientific question. These patterns serve as multi-property metaviews, \textit{e.g.} visual genotypes, enabling to classify material behavior and technique suitability according to predefined user demand and criteria, highlighting the interdependencies between measurement choices, extracted parameters and scientific interpretation. This multimodal workflow establishes a proof-of-concept for correlative analysis and underscores challenges toward fully integrated, automated and holistic approaches in energy material science.

\end{abstract}

% Text: Please use section headings and subheadings as specified below. For communications, all section headings apart from Experimental Section should be removed
% Please make the first reference to a display item bold: \textbf{Figure 1}
% Do not abbreviate Figure, Equation, etc.; display items are always singular, i.e., Figure 1 and 2.
% Equations are always singular, i.e., Equation 1 and 2, and should be inserted using the {equation} environment, not as graphics
% Please do not use footnotes in the text, additional information can be added to the Reference list.
%\twocolumn

\section{Introduction}

Batteries are complex electrochemical devices with many reaction and degradation processes spanning across extended length and time scales \cite{tarascon_issues_2001,fichtner_rechargrable_2022}.
Understanding the interdependency factors that control material and cell evolution is challenging due to the number of inter-related phenomena determining battery performance and durability - formation of fatigued phases, cracking of active particles, gas generation, solid electrolyte interphase growth, chemical cross-talk, microstructural changes, delaminations, corrosion, dendrite growth, lithium trapping and plating, \textit{etc.} \cite{vegge_toward_2021, narayan_self-healing_2022}.
Moreover, the primary causes of degradation and capacity loss depend on the type of active and inactive materials used, the way components and interfaces are engineered, cells manufactured and integrated, as well as the conditions of usage (voltage range, temperature, pressure, cycling rate, \textit{etc.}).
It is recognized that holistic and non-destructive multi-scale monitoring and diagnosis is one key to unlock system optimization and accelerate the discovery process \cite{atkins_accelerating_2022,amici_roadmap_2022}.

The quest to gather multidimensional arrays of parameters that describe battery state-of-charge (SoC) and state-of-health (SoH) relies on using a wide range of experimental techniques \cite{grey_sustainability_2017,li_peering_2021,zuo_nondestructive_2025}.
Not only multiple, heterogeneous, sets of data must be acquired but they should be correlatively analyzed to produce the desired result integration \cite{muller_multimodal_2020,lubke_origins_2024,dini_review_2021,ziesche_multi_2023,wolff_operando_2025,kramar_correlative_2025,bai_guidelines_2024,meng_multiscale_2024,ziesche_4d_2020}.

To achieve these goals, new infrastructures and methodologies are needed, as it is not feasible for a single expert to master all facets of battery characterization nor define universal experimental procedures; it instead necessitates the collaborative effort of many experts within an extended community. 
In addition, the community efforts cannot simply involve every field experts focusing independently on their specific area of research. 
Approaches where individuals or groups gather knowledge and information in parallel do not facilitate a consistent, final correlation of data, notably due to reproducibility issues, along with other factors such as a methodological bias \cite{puls_benchmarking_2024}. 
The lack of standards and quality control assessing the representativeness of results, their reliability and scalability, is also an issue \cite{frith_non-academic_2023, drnec_battery_2025}.
Hence, to design effective workflows for materials discovery and optimization, we need to foster cooperation, construct community tools, and bridge communities beyond single-site single-expert approaches, building on large scale collaborative initiatives such as Battery 2030+ in Europe \cite{amici_roadmap_2022,ahlgren_battery_2023}, and focused science hubs (like synchrotron battery hubs \cite{lyonnard_building_2025}).

Battery material acceleration platforms rely on experimental and numerical workflows (or workflow frameworks) to synthesize novel materials, characterize their properties, and model, predict and optimize their function in devices \cite{stier_materials_2024,benayad_high_2022,vogler_brokering_2023,schaarschmidt_workflow_2022, Vogler_autonomous_2024, steensen_necessity_2026}

Numerical workflows have been designed for the investigation of materials in a variety of fields \cite{friederich_toward_2019, mostaghimi_automated_2022, rego_simstack_2022, steensen_interoperability_2025}, demonstrating clear benefits in bridging research fields and simulation methods. 
Their general idea is to boost result acquisition by automatically interconnecting various steps, and to ease access to non expert users by encapsulating and automatically determining some of the model parametrization. For instance, Schaarschmidt \textit{et al.} \cite{schaarschmidt_workflow_2022} have reported workflow engineering for virtual material design, emphasizing non-expert friendly initiatives like the Pipeline Pilot \cite{warr_scientific_2012} used to automatically compute electrolyte transport properties from a selection of precise electrolyte compositions.
Additionally, the PerQueue workflow manager can be used to orchestrate high-throughput characterization data processing and simulation tasks, enabling consistent execution and provenance capture across sites \cite{sjolin_perqueue_2024}.

Experimentally, workflows are needed to improve streamlining and repeatability in the result acquisition \cite{puls_benchmarking_2024,drnec_battery_2025}, and to develop high-throughput autonomous materials discovery and screening. 
An increasing number of publications show the relevance of knowledge-informed, generalizable and hybridized methodologies, applied to problem-solving or chemical space exploration. 
For instance, the understanding of complex and interactive phenomena occurring during battery lifetime was rationalized by Meunier \textit{et al.} \cite{meunier_design_2023} introducing a method that combines electrochemical cycling protocols and advanced data analysis.
This workflow was showcased on selected Ni-rich chemistries, but enables screening other types of new electrolytes and battery chemistries based solely on the knowledge of electrochemical performances.
Using another strategy, optimization of battery electrolyte properties was also demonstrated by Vogler \textit{et al.} \cite{vogler_brokering_2023} using an international framework combining numerical and experimental studies with AI orchestration. 
Automatically orchestrated activities were ran across five international tenants over two weeks, leading to identifying key factors for success, result confidence and building more complex workflows.
Among these factors, agreements on data formats, limitations awareness, requests and answer traceability, ability to handle bad or missing data and a cost accumulation awareness for each tenant, were found to be critical. 
Regarding material developments, initiatives towards high-throughput material synthesis, property screening, cell assembly and testing, where robotized equipments are connected and controlled remotely with no, or little, human intervention, are emerging \cite{stein_progress_2019, sanin_integrating_2025}.
Automated battery labs and platforms using input characterization and modeling data are developing worldwide to improve on the usual trial-and-error strategies \cite{szymanski_autonomous_2023,fei_alabos_2024, zhang_ivoryos_2025, pablo-garcia_affordable_2025, battaglia_toward_2025}, as in other key sectors such as electronics \cite{strieth-kalthoff_delocalized_2024, he_algorithm-driven_2024}. 
These, however, do not aim at unraveling the origins of the observed behaviors, but rather optimize the way we search for enhanced properties.

Unlike for simulation or robotics, workflows aiming at advanced experimental characterization are less developed. 
Multi-technique and multi-modal approaches are recognized pivotal to obtain in-depth mechanistic understanding of battery behavior, but, in practice, attempts to develop correlative characterizations remain scarce and often technique-limited.
For instance, X-ray imaging methods \cite{withers_x-ray_2021} have been applied on battery materials since decades to access morphological and structural characteristics \cite{scharf_bridging_2022}, but only recently were extended capabilities and couplings demonstrated using hierarchical X-ray computed tomography (XCT) analysis.  Three-dimensional mapping from cell scale down to particle level was performed combining zoom-in zoom-out conditions, either on one instrument at different resolutions or using several complementary XCT instruments \cite{muller_multimodal_2020, scharf_bridging_2022, zan_understanding_2021, pietsch_x-ray_2017}. XCT was also combined with other imaging modalities such as neutron-based radiography or tomography, \cite{ziesche_4d_2020, lubke_origins_2024} or a diversity of lab-scale techniques \cite{zuo_nondestructive_2025, fordham_correlative_2023, du_-situ_2022, muller_multimodal_2025}.
Similar coupled approaches have been applied using electron microscopy techniques in the battery or material science fields, where the criticality of sample preparation and transfer methods have been pointed out \cite{liu_fib-sem_2025, vanpeene_sequential_2021, cadiou_multiscale_2020}. 
Generally, coupling battery imaging with complementary tools such as diffraction or sensing also enables to correlate changes in micro~/ nanostructure to local chemistry, gas formation, strain, pressure or temperature changes, \textit{etc.} \cite{nelson_operando_2012,olgo_revealing_2024}. 
In the same spirit, coupling or combining diffraction and spectroscopy techniques is required to link the chemical state of a redox center, for instance, to the lithiation degree of the host structure, providing possibilities to determine lithium inventory loss in real time during cell cycling, as obtained by coupling X-ray diffraction and fluorescence \cite{dawkins_mapping_2023} or X-ray absorption \cite{balasubramanian_situ_2001}.
Multi-spectroscopic workflows coupled with DFT are also suited to elucidate charge compensation mechanisms, as reported recently in LiNiO$_{2}$ (LNO) electrodes \cite{jacquet_fundamental_2024}. 
Finally, lab-scale apparatus and portable equipments can also be coupled to advanced synchrotron techniques to correlate specific parameters. 
For instance, On-line Electrochemical Mass Spectrometer (OEMS) coupled with scanning microdiffraction revealed how the gas generated in a battery is linked to the structural damage undergone by the cathode material during overcharge \cite{jacquet_mapping_2025}.
Nevertheless, existing multimodal and correlative studies are usually limited to few instruments and~/ or few conditions.
To the best of our knowledge, there is so far no demonstration of a performed large scale holistic experimental workflow for battery characterization.

Here, we fill this gap by reporting the concept and implementation of a multi-technique European-wide platform developed within the BIG-MAP project in Battery 2030+. 
We designed an extended experimental workflow unifying complex and high-end experimental techniques across multiple sites, multiple scales and a wide array of observables. 
The workflow application is showcased on a specific Li-ion chemistry, \textit{i.e.} Graphite~/ LNO system, cycled with either a standard or a modified carbonate-based electrolyte. 
Experiments were performed to detect and analyze signs of aging in standardized materials at the end of life, as well as to investigate the influence of the electrolyte on battery capacity fading mechanisms. 
After introducing the workflow concept and showing its practical implementation, we describe the workflow outcomes as a three-layered knowledge space: i) individual datasets provide specific mechanistic insights or single-property evaluations of the probed materials at a given scale with a given precision;  ii) all datasets integrated into global synthesized overviews, or metaviews, provide a discrete-type two-dimensional visualization of the results, enabling a false~/ true analysis of changes in the electrodes as a function of their history, measurement type and associated scientific question; and iii) correlated datasets, \textit{i.e.} leveraged results from a subset of techniques, bring specific knowledge into a chosen sub-area of investigation, depending on the material, criteria and goals.
This analysis reveals the complex interdependencies between scientific interpretation, results and methods. 
The metaview representations serve as graphical genotypes, encoding the interactions between knowledge and tools and their variability depending on the choice of materials, conditions and aims. 
We discuss the benefits, pitfalls, challenges and current limitations of this methodology, which is a first step towards a pan-European multimodal infrastructure deploying large scale coordinated workflows to advance our understanding of battery materials and devices.

\section{Workflow concept}

\begin{figure*}[htpb]
	\centering
	\includegraphics[width=0.95\linewidth]{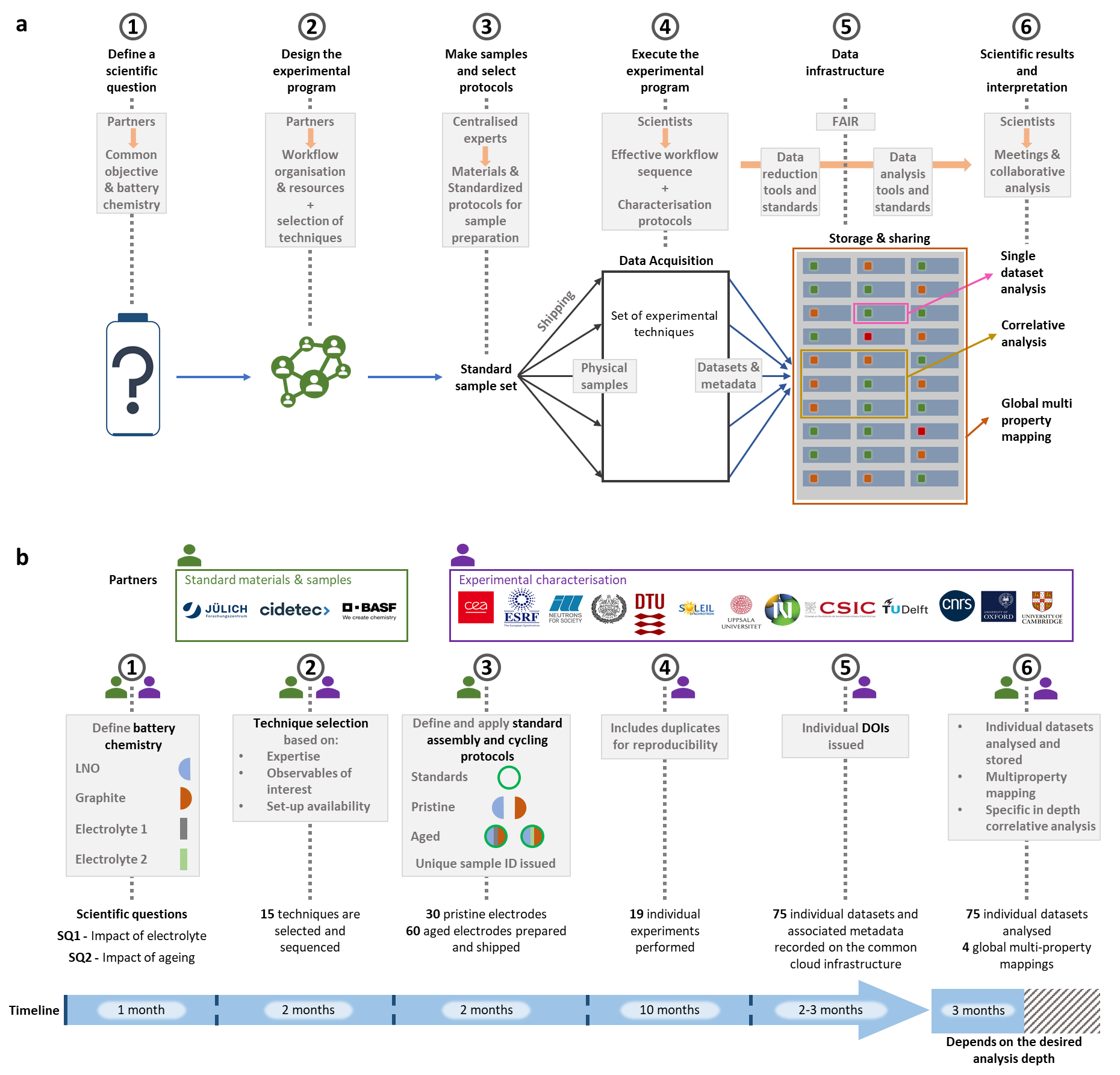}
	\caption{Concept and practical execution of a coordinated battery investigation workflow. a) Conceptual organization. Steps 1 to 3 involve preparation, step 4 corresponds to execution and steps 5 to 6 concern data storage, sharing and results exploitation. The analyses in step 6 can be divided into three categories: single dataset exploitation, global multi-property mapping, and correlative analysis. The light gray boxes and orange arrows indicate who is involved in each step and what specific actions are performed. b) Practical implementation of the workflow to study Gr~/ LNO batteries. Battery-making expert partners are indicated in green, characterization expert partners in purple. Time-sequenced execution of the 6 workflow steps, indicating who is involved (green~/ purple persons), what decisions are made and actions taken (gray boxes), and what are the main quantified outcomes (bottom text): 1) two scientific questions defined and battery components selected, including electrolyte 1 (standard) and electrolyte 2 (with additive); 2) selection of 15 relevant techniques; 3) production and shipping of 90 samples; 4) execution of 19 experiments; 5) 75 datasets and metadata handling through a common cloud infrastructure and 6) data analysis using three-layers: 75 individual datasets; 4 global multi-property metaviews and some correlatively-analyzed subsets. The timescale of the last analysis step has been limited to 3 months in our proof-of-concept application, but can be extended depending on the depth of targeted sub-analyses.}
	\label{fig_concept_workflow}
\end{figure*}

\textbf{Figure \ref{fig_concept_workflow}a} illustrates the conceptual building blocks of the workflow, which relies on the sequential execution of six main steps: 1) definition of a driving scientific case; 2) design of the adequate experimental program; 3) selection of samples and protocols; 4) execution of the experimental program (technique implementation and data generation); 5) data organization, storage and sharing among the workflow partners and 6) data analysis and correlated result integration.

This structure is reminiscent of the CHADA method \cite{romanos_innovative_2019} where key elements of the characterization chain are defined as sample, technique, probe and data categories. 
The core aim of the workflow is to provide a chemistry-neutral and versatile method to analyze the behavior of battery materials according to user-defined criteria and integrating user experience. 
It entails cooperation, discussion and organization to define its initial operative base (steps 1 and 2), while being a dynamic and modular infrastructure that embeds evolving functionalities by design (steps 3, 4 and 5), all steps being fully adjustable and tunable on-demand.
The programming and realization of the various subtasks along the six key steps necessitate a designed collaborative organization based on community tools and shared know-how. 
Following conceptual guidelines, as per the FAIR data principles, is instrumental to streamline collaborative data and metadata storage, sharing and analysis \cite{wilkinson_fair_2016}.

\section{Workflow implementation}
The workflow concept has been applied to a specific scientific case and collectively executed by 15 partners across Europe: CEA (France),  ESRF (France), ILL (France), FZJ (Germany), CIDETEC (Spain), CSIC (Spain), DTU (Denmark), Soleil (France), Chalmers (Sweden), UU (Sweden), NIC (Slovenia), TUD (Netherlands), BASF (Germany), UCAM (UK) and UOXF (UK) to showcase its capabilities, as summarized in \textbf{Figure \ref{fig_concept_workflow}b}.
This demonstration workflow was designed to perform the characterization of a chosen battery chemistry, employing a broad screening approach to explore various relevant aspects.
The workflow experimental part was ran across 10 months (excluding the design and preparation work) and actively gathered up to 35 researchers across all partners.
In the interest of time and sample integrity, it was mainly set in a parallel fashion and included organizational tools, standard protocols and storage~/ sharing capabilities.
After conducting the experiments, as well as during the workflow execution, ensuring a smooth process relied heavily on information sharing, encompassing both data and metadata.

\subsection{Scientific questions formulated}

The partners jointly decided to focus on two research topics of high relevance in today's battery research: advancing electrolyte optimization for liquid-type full cell batteries, and understanding aging mechanisms. A state-of-the art Li-ion battery was selected, using commercial grade materials, namely graphite at the negative electrode and LiNiO$_{2}$ (LNO) at the positive electrode. 
Graphite is the state-of-the art negative electrode material thanks to its high capacity \cite{zhao_progress_2024}, while LNO is the end-member of the layered oxide cathode materials NMC, having high capacity and cobalt free composition \cite{bianchini_there_2019}. 
At 80\% SoH, battery components usually exhibit altered morphologies, changes in electronic and crystalline structures, inactivated or damaged active particles, as well as presence of degradation layers such as the Solid Electrolyte and Cathode Electrolyte Interphases (SEI and CEI) \cite{kabir_degradation_2017}.
Therefore, to obtain a comprehensive overview of aged cells properties, bulk and surface techniques should be applied at all relevant scales to observe and quantify phenomena as diverse as lithium heterogeneities, surface modifications or strained particles.
Two wide range scientific questions (SQ) were identified as per step~1 of the workflow. 
SQ1: ``Is the addition of a property-building compound into the standard carbonate mixture used as electrolyte affecting the electrodes properties after long-term cycling?" and SQ2: ``Is long-term cycling affecting negative and positive electrode properties?"
These SQ can be answered by \textit{yes}, \textit{no}, \textit{maybe}, or \textit{I don't know}, but they also implicitly contain several extended forms, like: ``How does this happen?" or, ``Why does this happen?", which define another layer of details and targets, losing generalization but gaining complexity. 
Accordingly, in the first place, a SQ can be treated using a unique discrete variable (having values as true, false or null), leaving the possibility to further introduce sub-categories of discrete variables, metrics or quantities connected to information sub-layers, if additional knowledge is needed. 
SQ2 (aging), for instance, may be approached using a variety of intermediate questions to understand the precise nature of the degradation(s), their origin, their physical or chemical manifestations, spatial and temporal domains of existence, dependence on external factors, \textit{etc.}.
One could ask: ``Is the long-term cycling responsible for microstructure modifications in the negative electrode?"; ``What is the reduction in porosity measured at 80\% state-of-health?"; ``What is the link between overpotential and porosity?", \textit{etc.}.
These considerations indicate that the workflow can intrinsically capture multiple elements of an answer, which can be correlated within and across different knowledge layers. 
As such, it encodes a variety of combinatorial analysis or extendable formats, enabling answers and conclusions adjustable to selected user criteria and scopes. 
As our goal is not to cover all potentialities of the multimodal workflow, we focus on the representation of global answers to SQ1~/ SQ2, rather than fully exploring multi-correlative dimensions and reporting specific scientific results, which rather belong to specialized publications. For practical implementation, SQ1 and SQ2 will be formulated in a common form: SQ1 ``Is there a difference between samples prepared with different electrolytes?", and SQ2 ``Is there a difference between pristine and aged samples?", resulting in each researcher being able to interpret each of them as: ``When I look at several workflow samples with my technique, do I detect any difference between them?".

    \subsection{Samples, techniques and protocols}

The investigated materials were industry-grade LNO electrodes provided by BASF and Graphite electrodes from Cidetec (see Note 1 of the Supporting Information for characteristics). %\ref{Ssec-assembly}
The chosen workflow samples are labeled as ``aged LNO" and ``aged graphite". They were cycled in standard coin cell in full cell configuration (Figure S1, Supporting Information), using carbonate-based liquid electrolytes with or without LiTDI as an electrolyte additive, and dismounted at 80\% state-of-health (SoH). 
This SoH corresponds to the usual industry-grade end of life state (EoL). 
As cells were dismounted in a discharged state, the LNO electrode is lithiated and graphite delithiated.
The LiTDI additive was being investigated because of demonstrated improvement in cycle life for NMC811 cells and other Ni-rich materials \cite{berhaut_new_2019, niedzicki_new_2011, paillet_determination_2015, paillet_power_2015, pan_ethylene_2022, xu_litdi_2017} (see Note 1 and Figure S2 of the Supporting Information for more details). %\ref{Ssec-assembly} 

A combination of complementary techniques available and~/ or mastered by the partners for either ex situ, in situ or operando characterization, was chosen to uncover various aspects of chemical and physical evolution of the samples and achieve holistic views and mechanistic understanding (step 2 in Figure \ref{fig_concept_workflow}b). This technique selection step bridges steps 1 to 3, further enabling workflow execution (step 4).

\begin{figure*}[htpb]
	\centering
	\includegraphics[width=0.95\linewidth]{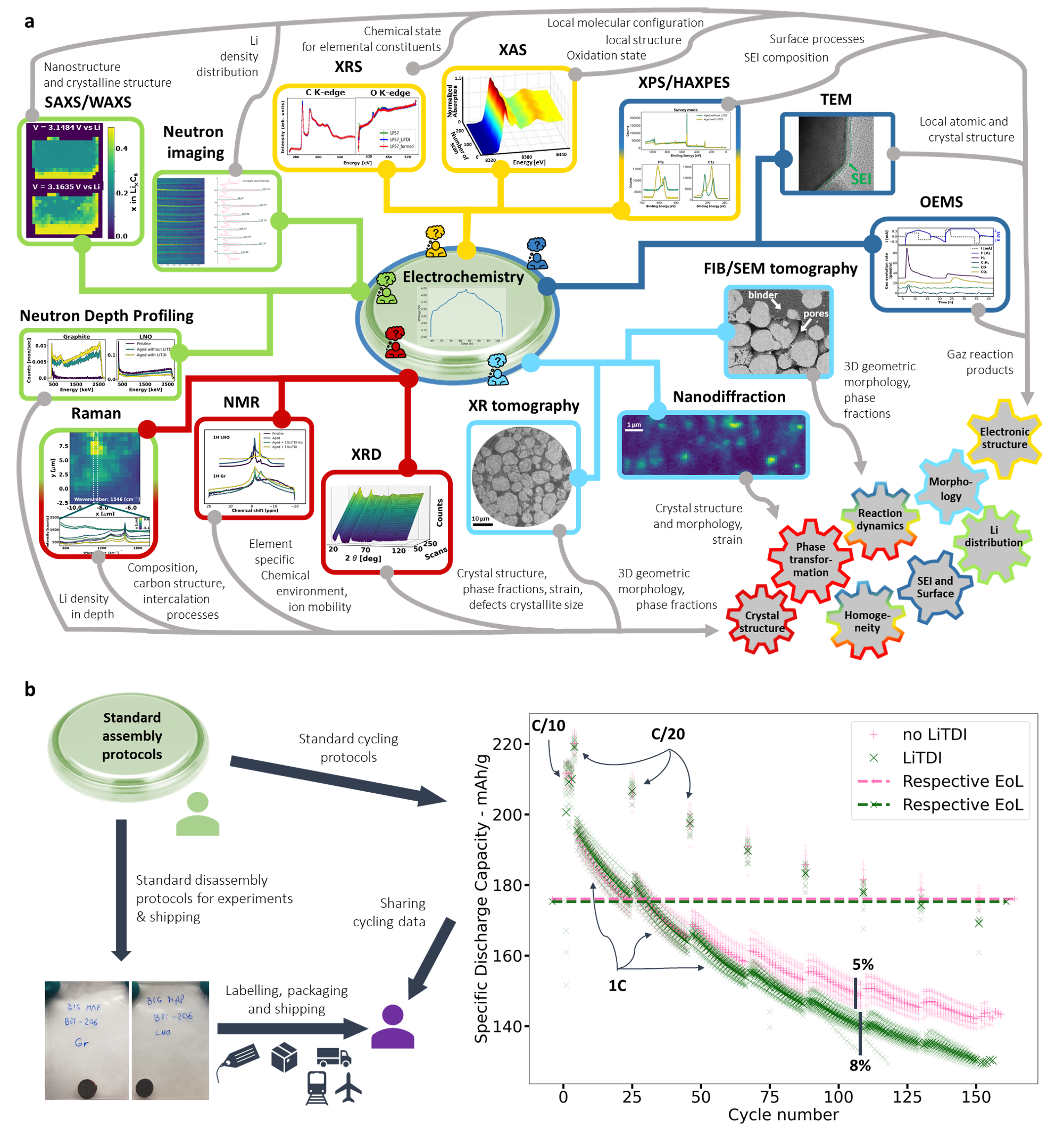}
	\caption{Techniques, observables and samples. a) Detailed view of workflow step 2 showing the combination of 15 techniques employed (colored boxes) and the physico-chemical parameters extracted (gray lines) to analyze crystal and electronic structures, phase transformation, morphology, Li distribution, reaction dynamics, homogeneity and SEI and surface properties. The experiments where performed on centrally standardized samples obtained from coin cell cycling. b) Execution of step 3 comprising the standardization of sample preparation and sample exchange between partners. Galvanostatic cycling data (discharge capacity \textit{vs.} cycle number) of 35 coin cells assembled with the two electrolyte types, showing 1C charge~/ discharge sequences followed by control measurement at C/20 to regularly evaluate the remaining capacity. The end-of-life is represented by the dashed lines, and corresponds to the cells being dismounted in discharged state to prepare the ``aged electrodes", \textit{e.g.} aged LNO in lithiated state and aged graphite in delithiated state. Electrolyte with LiTDI is plotted in green, and electrolyte without LiTDI in pink.}
	\label{fig_technique and sample}
\end{figure*}

\textbf{Figure \ref{fig_technique and sample}a} shows the 15 experimental techniques employed in the study, both in-lab: Electrochemical cycling, gas analysis (OEMS), Focused Ion Beam~/ Scanning Electron Microscopy (FIB~/ SEM), Nuclear Magnetic Resonance (NMR), Transmission Electron Microscopy (TEM), X-ray Diffraction (XRD), Raman 2D mapping, X-ray Photoelectron Spectroscopy (XPS), and at large scale facilities: two neutron-based techniques (Neutron Imaging (NI) and Neutron depth Profiling (NDP), performed in two different facilities, namely ILL and TUD) and 6 synchrotron-based techniques performed at the ESRF (X-ray nano holotomography (nanoCT), X-ray Raman Scattering (XRS), Small and Wide Angle Scattering (SAXS~/ WAXS), X-ray nanodiffraction microscopy (SXDM)) and SOLEIL (X-ray Absorption Spectroscopy (XAS), XPS~/ HAXPES).
The techniques can be grouped within 6 main categories depending on the scientific information they can provide: crystal structure (dark red) and phase transformation (red); electronic structure (yellow); SEI and surface properties (dark blue); lithium distribution (green); morphology (light blue), and two cross-categories:  reaction dynamics and homogeneity (multicolored by nature).

Once the techniques were selected, expert partners at Cidetec and J\"ulich were in charge of sample standardization, \textit{i.e.} preparing and providing the workflow samples further distributed to every partner in a timely coordinated manner.
To guarantee sample reproducibility, advanced standardized protocols were determined at different levels (\textit{cf.} \textbf{Figure \ref{fig_technique and sample}b}): i) cell assembly; ii) cell cycling and iii) cell disassembly.
Each of these standardized protocols are integrating preexisting internal recipes and expert know-how. 
Details on cell manipulation and testing standards can be found in Note 1 (cell assembly), Note 2 (standardized cycling protocol), Note 3 (electrochemical data with Figure S3) and Note 4 (cell disassembly) of the Supporting Information, and in \cite{cadiou_D5-3_2023}. %\ref{Ssec-assembly} - \ref{Ssec_disassembly}
Labeling, packaging and shipping procedures were normalized and executed by the same coordinating partner using the same methods.
These precautions insured a high degree of repeatability and traceability, minimizing errors and bias arising from the various key steps in the workflow.
The list of samples with their individual label and dates of delivery is provided in Note 5 (Supporting Information). %\ref{Ssec_stdList}. 

\subsection{Workflow execution}
    
In total, 74 coin cells were cycled by the two cell-making partners.
From these 74 coin cells, 35 were dismounted in discharged state to extract both the positive and negative electrodes probed during the workflow.
These selected cells exhibited a discharge capacity within 8~\% relative variation (see \textbf{Figure \ref{fig_technique and sample}b}).
The electrodes were collected, prepared (\textit{e.g.} cut, conditioned, handled) and distributed by the coordinating partners (ESRF, CEA).
The characterization experiments were further conducted either ex situ, in situ or in operando mode, depending on the constraints from each set-up and the targeted observables.
Including reproducibility checks from different partners, 19 individual experimental characterization actions were performed on 60 different standard electrodes (two electrodes per cell, one positive and one negative) for the aged samples and 30 pristine samples.
The acquired data are very diverse in size and shape: from 5 kilobytes 1D spectra from vibrational spectroscopy to 1 terabyte images collected by synchrotron nano holotomography, and so are the quantities extracted from them: lattice parameters, compositional disorder, band gap states, gas products, porous fractions, \textit{etc.}, as seen in Figure \ref{fig_technique and sample}a.
All these are key descriptors of the battery material status in a given SoC and SoH.
Their correlations are often not analyzed although they all describe reaction and degradation mechanisms. 
Based on the workflow task force and technical know-how, we defined the 8 main categories of interest for further result categorization and analysis, as already pointed earlier: morphology; crystal structure; lithium distribution; phase transformation; reaction dynamics; homogeneity; Solid Electrolyte Interphase (SEI) and interfaces and electronic structure. 

\subsection{Data infrastructure}

We addressed the challenge of FAIR-guided data and metadata sharing by utilizing the tools developed in the BIG-MAP project \cite{noauthor_bigmap_nodate}, such as the BIG-MAP Archive (\url{https://archive-capex.energy.dtu.dk/}) \cite{granata_implementing_2025} and Notebook which has been made redeployable beyond BIG-MAP through the BIG-MAP Archive API client on Zenodo (v1.1.0, 11184798).
Such tools, along with the tracking of individual samples through simple sample identifiers (\textit{cf.} step 5 in Figure \ref{fig_concept_workflow}b), facilitated the effective information sharing throughout the workflow. Accordingly, each partner filled the standardized Notebook forms to declare metadata and stored the raw data in the Archive. 
One experimental campaign resulted potentially in several entries in both Notebook and Archive, as the storing is split at the level of individual samples. 
For some large-scale facility data, a full data transfer was not performed due to the very large volumes and to avoid unnecessary duplication, as these datasets are already securely stored and curated at the facilities. 
Instead, the facility-assigned DOI was used to reference the raw data and only processed or analyzed sub-datasets were uploaded to the shared platforms.
These procedures are required for a future successful result exploitation from the workflow both within the workflow partners, the project consortium, and the entire battery community when the data will become open access as per request of the BIG-MAP data management plan \cite{castelli_data_2021}.

\section{Data and results exploitation}

Each dataset of the workflow was analyzed first as an individual dataset by each experimental lead partner, using its usual data analysis pipelines and methods (which generally do not integrate other experimental inputs nor correlative analysis), and then as a piece of the ensemble data, \textit{i.e.} introducing global representations gathering all datasets into holistic metaviews. In the following, we first briefly describe the value of individual datasets and their potential meaning for materials analysis. Then, we develop the metaviews constituting the core of the discrete-type workflow output. Last, we show how intermediate specific knowledge can be gathered by selecting sub-groups of analyzed datasets. Overall, the analysis process produces a three-layers result space, from individual stand-alone results to global property mappings, with access to specific sublayered (partially correlated) results.

\subsection{Individual datasets}\label{sec_indivDatsets}

A total of 75 datasets were generated in the workflow. Details of each individual datasets acquisition and qualitative analysis are described in the Supporting Information for lab-scale techniques (Figure S3 to S18, Note 3 and 6 to 12, Supporting Information), synchrotron-based techniques (Figure S18 to S25, Note 12 to 17, Supporting Information) and neutron techniques (Figure S26 to S29, Note 18 and 19, Supporting Information). 
As said earlier, the 15 experimental techniques available in the workflow provide information on globally 8 categories of parameters. %\\~\\

\threesubsection{Reaction dynamics} 
The graphite~/ LNO cells exhibit some continuously degraded electrochemical performances along long-term cycling at 1C, as seen from the cycling data of coin cells used to create the aged workflow samples (\textit{cf.} Figure \ref{fig_technique and sample}b and Note 3, figure S3 in the Supporting Information). All cells present, as expected, some capacity loss during the check-up cycles performed at C/20 when compared to the first formation cycles. %\ref{Ssec-cycle}
Interestingly, this capacity loss rate is comparable for both electrolyte formulations. 
Therefore, no cycle life improvement is witnessed using LiTDI additive with LNO, in contrast with NMC. 
However, operando gas analysis (OEMS) (\textit{cf.} Figure S6, Supporting Information) on a cell aged without LiTDI reveals that the onset of CO$_{2}$ emission in the LNO electrode at high states of charge (above 4.2 V) is lower than that of typical high Ni-content materials as NMC. % \ref{Ssfig_OEMS}
This is linked to degradation mechanisms such as lattice oxygen loss or a rock salt layer formation on the active material surface, hence degradations are expected in LNO electrodes while cycling with a cut-off voltage above 4.2 V.
In tune with CO$_{2}$ emission, a cross talk with the Gr anode is witnessed through H$_{2}$ gas emission linked to VC transformation and SEI formation. 
Therefore, it seems that some microscopic-level changes are induced by the presence of the additive with respect to the benchmark system using only standard carbonate mixture and LiPF$_{6}$ salt, although electrochemical data at cell level are similar. This observation highlights the need for more in-depth investigations at different scales to complement electrochemical cell cycling analysis.\\
%\\~\\

\threesubsection{Crystal structure~/ phase transformations} 
LNO is known to undergo a series of phase transitions during lithiation~/ delithiation, where the \textit{a} and \textit{c} parameters of the layered oxide structure vary in function of the lithiation content, \textit{x} in Li$_{x}$NiO$_{2}$ \cite{bianchini_there_2019,xu_phase_2020, jacquet_mapping_2025}.
Particularly, the collapse of the structure at potentials higher than 4V is accompanied by anisotropic strain development, believed to generate cracks and loss of active material during repeated cycles \cite{jousseaume_strain_2024}. %(cite Jousseaume et al, EES 2024). 
In the workflow, laboratory X-ray diffraction was performed both operando and ex situ on the aged LNO electrodes cycled with and without LiTDI (\textit{cf.} Note 10, Supporting Information). %\ref{Ssec-xrd}
It was concluded that the crystalline structure, unit cell parameters and phases remain similar to the pristine material. 
Nevertheless, the two types of aged LNO show distinct residual charge states, with observed differences in the mix between rombohedral (lithiated) and monoclinic (less lithiated) phases. 
This result suggests that cycling with LiTDI induces some lithiation heterogeneities on the long-term.
Note that complementary operando synchrotron SAXS~/ WAXS experiments were conducted on the full cell aged system (LNO \textit{vs.} Gr) to provide additional structural details (see Note 15, Supporting Information). %\ref{Ssec-saxs-waxs}
However, due to a labeling or communication error, the positive electrode material was NMC rather than LNO, rendering the results irrelevant to the present workflow. 
Nevertheless, the experiment was kept in the experimental entry list, as it produced data and highlights the importance of traceability and reliability in experimental workflows.

\threesubsection{Electronic structure}
Monitoring the electronic states of the redox centers and local chemical environment is key to understand the charge compensation mechanisms which accompany the insertion and deinsertion of lithium ions in~/ from the host structures. 
XAS investigated the difference between pristine (discharged) and aged LNO electrodes with LiTDI  (\textit{cf.} Figure S24, Supporting Information). %\ref{Ssfig_XAS}
The Ni-O bond is shorter in the aged LNO indicating a higher Ni oxidation state. 
Additionally, the intensity of the Ni-O bound peak is higher in the aged electrode compared to the pristine state, maybe due to a higher ordering of the NiO$_{6}$ octahedrons with cycling. %\\~\\

\threesubsection{SEI and surface processes} 
Controlling the SEI structure and chemistry is of utmost importance for long lasting batteries. 
Tuning the electrolyte composition, and achieving synergistic effects using one or several additives, is a widely-used strategy to build robust and stable SEIs in a variety of systems \cite{sinha_storage_2013,zheng_critical_2021}.

However, detecting and quantifying SEI species and morphology remains a challenge due to the difficulty to access interfacial layers buried in the composite materials. Additionally, these interfacial layers are very sensitive to sample manipulation, potentially unstable in time and highly heterogeneous in size and nature \cite{adenusi_lithium_2023}. 
The TEM observation of the SEI on model samples reveals that samples cycled without LiTDI exhibit a uniform layer at around 2 to 3.5~nm, while samples cycled with LiTDI have a less uniform SEI with a thickness that can vary from below 1~nm up to 6~nm within one particle (\textit{cf.} Figure S12, Supporting Information). %\ref{Sfig_TEM_results_composite}
Note that the electrochemical cycling protocol to create these model SEI samples is similar to the one applied for standard cells used in the other experiments.

$^{7}$Li and $^{19}$F solid state NMR was performed ex situ on both positive and negative electrodes. The $^{7}$Li peak width (\textit{cf.} Figure S9a, Supporting Information) in the Gr electrode is broader in samples with LiTDI in comparison to samples without LiTDI, with a tail towards lower frequencies indicating a greater variety in the SEI environment and a more disordered structure. %\ref{Ssfig_NMR_Gr}
In the LNO electrodes, the Li$_{2}$CO$_{3}$ peak is more intense for the samples cycled with LiTDI (\textit{cf.} Figure S10a, Supporting Information). %\ref{Ssfig_NMR_LNO}
Similarly, the $^{19}$F spectra display PF$_{6}^{-}$ and LiF peaks both with and without LiTDI, the samples cycled with LiTDI having broader peaks indicating a more disordered SEI environment or a closer Gr surface  (\textit{cf.} Figure S10b, Supporting Information). % \ref{Ssfig_NMR_LNO}
The LiF peak is also slightly more intense in the samples cycled without LiTDI. These results indicate a distinct chemical composition of SEI when using LiTDI, which is compatible with the different morphology found with TEM.

XRS was performed on the aged Gr electrodes, bringing information on averaged chemical composition in the cycled samples. Analysis of the C and O K-edges respectively shows no influence of the cycling conditions on the local graphite atomic and electronic structure as well as no changes in the Li$_{2}$CO$_{3}$ content in the SEI (\textit{cf.} Figure S25, Supporting Information).
Note that the Ni and Li K-edge signals were non-informative.
The conclusion is that no significant changes were detected by XRS between the Gr electrodes cycled with and without LiTDI.

A key technique to investigate SEI is XPS, which is widely used in battery research to determine degradation species and quantify their concentrations \cite{shutthanandan_application_2019,kallquist_advances_2022}. 
The main limits of the technique come from the limited penetration depth and vacuum-constraint, which induce rinsing protocols and surface-limited information, potentially biasing the results.
Here, XPS was employed to investigate the SEI composition in both LNO and Gr samples aged with and without LiTDI.
The survey spectra exhibit matching patterns with more peaks in the sample with LiTDI (\textit{cf.} Figure S18, Supporting Information). %\ref{Ssfig_XPS_survey}
This indicates a different and more chemically complex SEI in the samples aged with LiTDI, in line with NMR results.
Mechanistic details obtained from the high resolution spectra analysis can be found in a dedicated paper \cite{herrera_standardization_nodate}. %\\~\\

\threesubsection{Morphology} 
Electrode microstructure is a key factor governing the diffusion pathways, reaction kinetics and charge transfers at the particle interfaces, which are known to be affected by electrode composition, active material sizes and nature as well as porous network characteristics \cite{zhou_engineering_2024}. 
Synchrotron nano holotomography is a powerful technique to access the microstructure at ultimate scales (\textit{e.g.} with nanoscale resolution) \cite{vanpeene_comparative_2025}. 
Because an electrode microstructure is complex, hierarchical and heterogeneous, advanced image processing methods are often required for quantitative analysis possibly supported by recent advances in deep learning \cite{rieger_utilizing_2024}. %(ref Laura). 
A standard approach was applied here to analyze the 3D morphology of both Gr and LNO electrodes in the pristine and aged states, with and without LiTDI (\textit{cf.} \textbf{Figure \ref{fig_science_summary_all}a} and Figure S19, S20, Supporting Information), revealing differences between aged and pristine LNO and Graphite electrodes, depending if the LiTDI additive is used or not (see Note S13, Supporting Information).

In Graphite, the phase volume fractions are estimated using a two-phase segmentation (dense particles $+$ low attenuation binder~/ SEI~/ void phase), showing a decrease in porosity 
in aged electrodes compared to pristine. Moreover, the electrode aged with LiTDI is more porous ($\sim$33\%) than the one cycled without LiTDI ($\sim$26\%), which could indicate more pore clogging due to more~/ thicker SEI building. 
However, these values must be taken with care, due to measurement uncertainty arising from lack of sample reproducibility and~/ or lack of statistical analysis.
Regarding LNO electrodes, there is a clear decrease in porosity \& binder volume fraction (both being indistinguishable) between the pristine state ($\sim$59\%) and the aged state, with the electrode cycled without LiTDI having a higher porosity ($\sim$48\%) than the one aged with LiTDI ($\sim$41\%).
This could be the signature of different particle reorganizations during cycling for the LNO electrodes.
Globally, (in)homogeneity assessment (yes~/ no answers to SQ1 and SQ2) is performed but within a given limited framework: only one piece of electrode was measured in most cases, analyzed with two-phase segmentation not at the highest resolution possible. 
Hence, the answers could vary if more complex quantitative image analysis was conducted at a smaller scale using higher spatial resolution, as performed in dedicated studies \cite{vanpeene_4d_2025, muller_multimodal_2020,liu_fib-sem_2025}. 

A technique complementary to X-ray tomography is FIB~/ SEM, which was also performed on LNO electrodes aged with and without LiTDI to challenge~/ complement the nanoCT results. 
It revealed an heterogeneous distribution of the conductive binder on the large scale cross sections (\textit{cf.} Figure S7, Supporting Information) and potentially also in the high resolution volumes (\textit{cf.} Figure S8, Supporting Information). 
Otherwise, there is no clear variations in the morphology between LiTDI and No-LiTDI LNO electrodes at the probed scale.

\threesubsection{Homogeneity and lithium distribution} 
All changes induced by aging and electrolyte formulations can be highly localized and heterogeneous, including lithium concentration - an effect that is not captured in the previous datasets, as the associated techniques all provide electrode-averaged ensemble results. 
Scanning or imaging techniques can overcome this by mapping properties in-plane or through-plane.
Particularly, lithium distribution in active materials is known to be heterogeneous after cycling, due to phenomena as lithium trapping, plating, thick SEI, particles disconnections, \textit{etc.} \cite{fang_quantifying_2019}. Lithium inventory loss is one of the major causes of capacity fade in batteries, together with loss of active material and polarization \cite{oney_dead_2025}. 

2D Raman mapping was used to analyze reaction heterogeneities, probing the electrode particles facing the separator in the pristine and cycled cells \cite{jacquet_fundamental_2024, flores_in-situ_2018}. 
In the LNO electrodes  (\textit{cf.} Figure S17a, Supporting Information), all states, from pristine to both aged samples, show similarly slightly heterogeneous structures linked to the LNO fully lithiated state (ratios between Ni$^{2+}$~/ Ni$^{3+}$~/ Ni$^{4+}$) both inter and intra-particles. %\ref{Ssfig_raman}
But globally there is no clear differences in Li distribution, crystal structures and state of charge in between the different samples.
In the Gr electrodes  (\textit{cf.} Figure S17b, Supporting Information), no intra particle heterogeneity was witnessed in all three electrodes, pristine, aged with and without LiTDi. %\ref{Ssfig_raman}
However, pristine and aged states show different I$_{D}$~/ I$_{G}$ band ratios, \textit{e.g.} 0.54 in pristine against 0.43 at 80\% SoH with LiTDI and 0.24 at 80\% SoH without LiTDI. Higher band difference indicates disordered graphitic or conductive additive domains.
Therefore, a degradation of the conductive pathways is observed in the cycled Graphite electrodes, with more degradation in the samples aged without LiTDI. 
This could be paralleled with the decreased porosity seen by nanoCT that may cause an increase in geometric tortuosity and hinder ion fluxes through the 3D electrode volume.

In addition, several techniques were employed to map lithium concentration at relevant scales, from particle to cell component level. 
Synchrotron X-ray nanodiffraction microscopy (SXDM) \cite{colalongoImagingInterIntraparticle2024, martensDefectsNanostrainGradients2023, wang_x-ray_2025} was used to investigate the strain and lithiation heterogeneities in the aged LNO crystallites (single crystals within the secondary particles). 
The measurement was performed in operando mode, coupled with bulk microdiffraction acquisition performed in between the SXDM acquisitions to monitor the overall electrode evolution using a larger beam. 
The time-resolved microdiffraction data showed expected evolutions of the diffraction patterns, \textit{e.g.} shift of the (003) Bragg peak showing the expected variations of \textit{c} parameter in LNO. 
Despite this, the particle-level SXDM acquisitions showed invariant diffraction patterns along cycling, revealing a strong beam effect locally, which impeded the normal electrochemical reaction to occur in the illuminated particle (\textit{cf.} Note 14, Supporting Information). Consequently, no particle-level information was obtained.  %\ref{Ssec-sxdm}

The full electrodes in the pristine and aged states, with and without LiTDI, were imaged ex-situ using Neutron Imaging in tomography mode, a 2D technique very sensitive to lithium due to its high absorbing cross section with respect to X-rays \cite{ziesche_neutron_2022, senyshyn_spatially_2014}.
Some qualitative changes were found by comparing the averaged attenuated intensities measured across the various elements.
In the aged LNO samples, values are comparable with and without LiTDI ($\sim$123 for both), lower than in the pristine ($\sim$127) (\textit{cf.} Figure~S27, Supporting Information). %\ref{Ssfig_NI_profile}
In contrast, aged graphite electrodes exhibit averaged intensities slightly higher than in the pristine state ($\sim$103 against $\sim$98, \textit{i.e.} a change by 5\%).
These results might indicate a loss in Li retention in the LNO electrodes, and more non-reactive Li trapped in the Gr electrodes (either inside inactive particles, SEI or both) with cycling. 
The similarities between cells aged with and without LiTDI is coherent with the electrochemistry showing similar retention capacities. 
Note that operando measurements or higher resolution tomographic acquisitions could bring more insights and enable a more precise localization and quantification of lithium content \cite{ziesche_neutron_2022,bradbury_visualizing_2023}, but they are costly in time and were not implemented in this workflow. 

Alternatively, lithium distribution within the different electrodes was analyzed by Neutron Depth Profiling, a technique suited to depth-resolve concentration gradients \cite{lv_operando_2018}. 
In coherence with NI, it has been evidenced that the aged Gr and LNO electrodes exhibit respectively levels of trapped Li and Li loss that are not present in the pristine samples (\textit{cf.} Figure S28 and S29, Supporting Information).
However, while NI showed similar levels in between samples aged with the different electrolytes, NDP reveals that the LiTDI electrolyte seems to amplify both the Li trapping and Li loss in the Gr and LNO, respectively, when comparing to the electrodes aged without LiTDI.
This could indicate that the additive does not promote more reversible reaction mechanisms and induce more lithium loss, potentially in the SEI.\\~\\

To conclude, the individual datasets reveal details on morphological features, SEI nature and composition, crystal structures, electronic states, and Li distribution for both chemistries with and without LiTDI. 
Clearly, the investigated LiTDI-containing chemistry does not hold promises to improve LNO~/ Gr battery performance for application purposes, but our multidimensionnal analysis raises interesting questions regarding the fundamental understanding of the physico-chemical behavior in batteries.
Indeed, while LiTDI and No-LiTDi cells display similar capacity retention and electrochemical performances until end-of-life, the analysis of the results obtained by other characterization techniques show significant differences in some of the electrode properties.
Hence, two different sets of electrode properties (including different SEIs) can induce the same overall electrochemical behavior. 
As a consequence, optimizing batteries can not only rely on monitoring electrochemical parameters as stand-alone health indicators, but rather necessitates closer insights into local microscopic-scale phenomena that compete and interfere. \\

\subsection{Property-mapping Metaviews}\label{sec_propMapppings}

\begin{figure*}[htpb]
    	\centering
    	\includegraphics[width=0.85\textwidth]{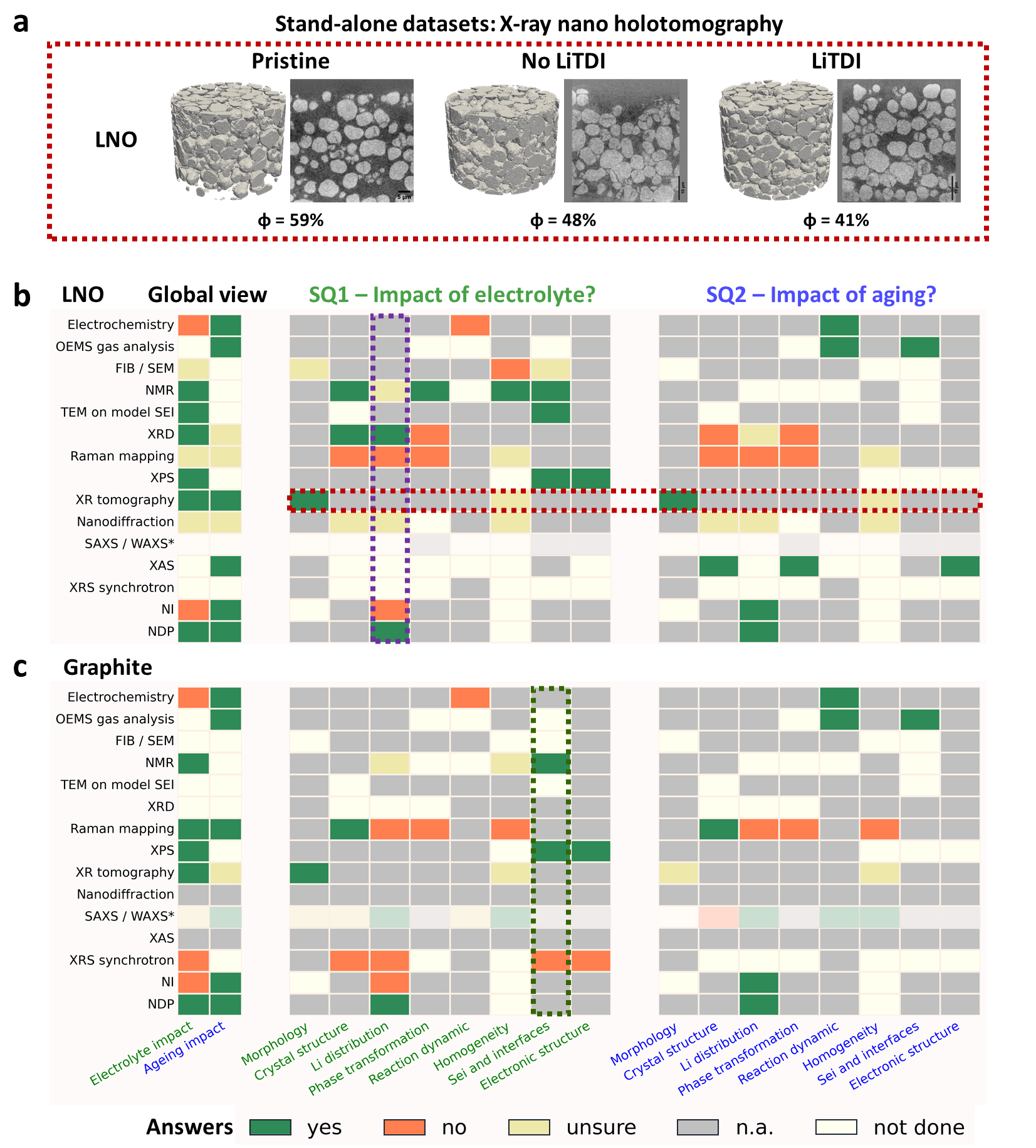}
    	\caption{Isolated dataset and result classification using global multi-property 2D metaviews. a) Stand-alone result: X-ray nano holotomography of LNO electrodes. 3D reconstructed volumes are shown for pristine as well as aged with and without LiTDI samples at high resolution with the corresponding estimated porosity ($\upphi$). The technique enables to track morphological evolutions during cycling with the two electrolytes (SQ1 and SQ2). b) LNO metaviews. c) Graphite metaviews. A metaview is a (15, 8) matrix (15 techniques, 8 observables) reporting discrete answers to SQ1~/ SQ2 (middle and right panels, respectively) into colored pixels (yes = green; no = orange; unsure = yellow; non applicable = gray; not done = uncolored). SQ1: ``Is there a difference between samples prepared with different electrolytes?" and SQ2: ``Is there a difference between pristine and EoL samples?". Global views (left panel) are composed by averaging the parameter-specific answers using a  prioritized logical rule sequence: i) one green = green; ii) one yellow = yellow; iii) one orange = orange; \textit{etc.}. Dashed rectangles highligth specific examples of correlated subset of results, picking one given technique (red applied to LNO for SQ1 and SQ2) or one particular parameter across all techniques (green for graphite, purple for LNO, applied to SQ1).}
    	\label{fig_science_summary_all}
    \end{figure*}

\threesubsection{Building global correlative representations} 
We define metaviews as two-dimensional space diagnosis maps or global property-mapping patterns formed using the 15 techniques (vertical axis) \textit{vs.} the 8 defined categories of observables (horizontal axis). Each intercept pixel (v, h) contains the answers to the defined scientific questions SQ1~/ SQ2 using a color code (\textbf{Figure \ref{fig_science_summary_all}b and c}). 
Practically, to build the metaviews for LNO and graphite, a survey was sent to the experimentalists by the workflow coordinators to collect the first conclusions drawn from their experiments and provide discrete answers (yes~/ no~/ unsure~/ n.a.~/ not done). 
The many answers were further gathered and assembled in the matrix-type representation resulting in pixelised colored patterns, as shown in Figure \ref{fig_science_summary_all}b and c for LNO and graphite, respectively. 
A positive answer comes as a green-colored box, while a negative answer is colored in orange, unsure in yellow, non applicable (n.a.) in gray, and not done is left as an uncolored box (white). 
The diagnosis maps regarding SQ1: ``Is there a difference between samples prepared with different electrolytes?" (green text) are shown on the middle panels of Figure \ref{fig_science_summary_all}b-c, while the results regarding SQ2 (blue text): ``Is there a difference between pristine and EoL samples?" are shown on the right panels. 
The left panel reports global views, as described later.
Looking into the patterns closely, we can identify several interesting features arising from the distribution and population of pixels. %\\

\threesubsection{Material identity map} 
Each pixel is given a color and hence visually contains the answer to the general overarching question using a specific tool and focusing on a specific aspect. For example, in map ``LNO - Global view", column ``Electrolyte impact” and line ``XPS”, the pixel is green, which means that LNO electrodes cycled with and without LiTDI exhibited significantly different XPS spectra.
Of course, the nature of this difference and further mechanistic understanding are not explicited in the metaview level analysis. Detailed exploration of the XPS stand-alone dataset must be performed to fully elucidate the observed phenomena - and the same applies for all pixels.

For a given material, the colored pattern changes when the question is different (Figure \ref{fig_science_summary_all}b \textit{vs.} c). 
Reciprocally, the same question investigated on graphite or LNO produces distinct 2D patterns, showing the specificity of each active material and highlighting their distinct response to electrolyte change or aging. Therefore, the metaview representation enables to collect visually the ``identity" of a material with respect to an associated question.
Extrapolating this, metaview patterns can be seen as a 2D barcode, a ``DNA-type" signature of sorts, each pixel being seen as a clickable gate towards stand-alone in-depth analysis (\textit{e.g.} a specific gene). %\\

\threesubsection{Missing, non-applicable or uncertain answers} 
We observe that a majority of metaview pixels are gray (non-applicable, n.a.) but that all columns or rows contain at least one colored box, indicating that the 2D space was consistently sampled in the present study. A gray box, by construction, indicates a mismatch between a given technique (for instance, ``Electrochemistry", row 1) and a given category (for instance, ``Morphology" or ``Crystal structure", columns 1 and 2, respectively). 
This simply tells that some techniques cannot provide relevant information about certain processes, mechanisms or properties. 
For example, the electrochemical cycling method used here (galvanostatic cycling) is informing about global reaction dynamics (only box (1, 5) can be colored), while X-ray tomography can provide morphology and homogeneity insights (boxes (9, 1) and (9, 6) are colored if the measurement was done), but no direct information about phase transformations or electronic structures (boxes (9, 4) and (9, 8) are gray whatever the sample or the scientific question).

Not-done pixels are also numerous in the patterns (shown in white). Missing characterizations are due to a conjunction of factors: lack of time to perform more experiments, lack of instrument availability, unsuited instrument, unavailable set-up for battery characterization, or decision not to implement additional experiments in order to keep the demonstrator workflow to a reasonable size. 

Finally, ``unsure” pixels are also spread within the patterns, corresponding to experiment applicable and performed, but yielding inconclusive results. 
This can be due to human errors (as for samples exchanged in SAXS~/ WAXS), technical issues (as beam damage for nanodiffraction), or challenging and time-consuming quantitative data analysis (as for ex situ X-ray nano holotomography in some conditions). %\\

\threesubsection{Yes and no answers} 
The large majority of green pixels suggests that most of the chosen techniques are sensitive to electrode changes during cycling, confirming the good choice of techniques. 
Nevertheless, certain techniques did not reveal differences induced by cycling electrodes with LiTDI (\textit{vs.} without LiTDI). 
This can be readily explained by the fact that the used techniques are probing different length scales or chemical~/ physical properties. 
Some experiments are also performed ex situ and others operando, which adds another potential difference. %\\

\threesubsection{Technique-by-technique visual analysis} 
Each metaview contains explicit information about technique sensitivity, applicability, and reliability by inspecting the sequence of colors in the rows of the matrix. 
For instance, XPS informs on SEI and surface properties as well as electronic structure, potentially overlapping in this regard with XAS, NMR, TEM, XRS, OEMS, but is not relevant to probe crystal structure or morphology. 
Moreover, XPS is a surface technique, therefore not probing the same samples areas as bulk ones (note that this information is not encoded in the matrix as such and requires prior knowledge from users for matrix interpretation). 
Its areas of relevance are located in columns 7 and 8, with potential extrapolation to column 6 (``Homogeneity"), all others being gray (n.a.). 
Alternatively, some techniques such as XAS or NMR have a wide spectrum of relevance, basically probing all parameters, except morphology and SEI for XAS, morphology and electronic structure for NMR. 
As a matter of fact, these techniques can provide sometimes yes and sometimes no answers, depending on the parameter of interest, revealing their adjustable sensitivity to various aspects of materials properties. 

Moreover, the overall technique response to a SQ can be assessed by ``combining" all related pixel information into one single colored pixel.
This result is gathered in the ``Global view" section on the extreme left of Figure \ref{fig_science_summary_all}b and c.
Such a combination follows a logical rule sequence with each rule having priority over subsequent ones: i) if there is at least one ``yes", then ``yes"; ii) if there is at least one ``unsure", then ``unsure"; iii) if there is least one ``no", then ``no"; iv) if there is at least one ``not done" then ``not done" and finally v) if there is only ``n.a.", then ``n.a.".
For example, in the case of ``Raman mapping" applied to Graphite for SQ1, the corresponding line, in the middle panel of Figure \ref{fig_science_summary_all}c, has at least one ``yes" pixel, then the pixel for ``electrolyte impact" in the ``Global view" panel (on the left) in row ``Raman mapping" is green.
Similarly, no data is reported on LNO by XRS for SQ2, leading to the presence of ``not done" and ``n.a." pixels only in line ``XRS synchrotron" for the right panel in Figure \ref{fig_science_summary_all}b, hence the ``Global view" pixel, left panel, for the corresponding line and column ``Ageing impact" is white.

As such, the metaview hence provides a quality assessment of the results including suitability factors, technique by technique. 
This shows that the metaviews can also serve as a basis to choose techniques and methods for further material evaluation or novel chemistries, using prior knowledge on other similar systems. %\\

\threesubsection{Parameter-by-parameter visual analysis} 
The patterns are more or less heterogeneous, depending on the parameter chosen to define the output result. 
Indeed, some categories (\textit{i.e.} columns) exhibit clear and robust answers (all pixels are of the same color), others remain uncertain (some pixels are unsure despite all others displaying the same color) and some reveal diverging answers (a mix of yes and no, typically). If one looks at LNO, SQ2, and ``Reaction dynamics", it seems that aging induces measurable changes, as found by both Electrochemistry and Gas analysis. 
This is a ``clear answer" situation, even if only two techniques were employed and therefore the result might be less representative than others.

In contrast, for instance, the case of Graphite $+$ SQ1 $+$ ``SEI and interfaces" (indicated by dashed green rectangle in Figure \ref{fig_science_summary_all}c) shows conflicting answers: NMR and XPS detect differences between the LiTDI~/ No-LiTDI samples but XRS does not (see Note 17 and Figure S25, Supporting Information).
An example of an uncertain situation is found regarding LNO, SQ1 and ``Lithium distribution" (dashed purple rectangle in Figure \ref{fig_science_summary_all}b), as the electrodes cycled with and without LiTDI show no distinct behavior using Raman mapping and Neutron Imaging, while XRD and Neutron Depth Profiling detected distinct signatures when using, or not, LiTDI (see Note 10, Note 19, Figure S14 and S28, Supporting Information, respectively) and NMR and nanodiffraction are yielding uncertain results, rendering the global assessment uncertain. 

The parameter-averaged assessment of one material behavior with respect to SQ1 or SQ2 can be extracted using the same logical rule sequence as the one used to compile the ``Global view" panels.
As seen from the global views, aging has most of the time a detectable effect (there is a majority of green pixels and no orange pixel in the corresponding column), while changing the electrolyte provides a more nuanced situation (a mix of orange, green, and yellow pixels). 
This reflects, on the one hand, the difficulty to access the electrolyte effect as it can be subtle or beyond reach with current experimental capabilities (in terms of resolutions, for instance, with the need to probe interfaces in real conditions) and, on the other hand, the potential interplay of beneficial and detrimental consequences induced by the additive, depending on what electrode properties is scrutinized. %\\

\threesubsection{Towards visual genotypes} 
Generally, differences were found with most techniques (```yes", green color) while some techniques gave an ``unsure” answer (gray) and a few techniques found no differences between LiTDI~/ No LiTDI and~/ or pristine~/ aged samples (orange).
The complexity of answering SQ1 and SQ2, that depends on the electrode material, can be grasped immediately from the visual rendering of the patterns, serving as first bricks towards material genotypes or IDs. 
Of course, critical information as surface~/ bulk capabilities, ex-situ~/ or in-situ~/ or operando modes, scales of observations, sensitivity or detection limits, \textit{etc.} are not directly encoded in the metaviews and require additional expert knowledge. 
But they could be added using more complex matrix representations where the user could select functionalities to focus on specific aspects. 
Looking forward, building machine-actionable, interactive databases of these material IDs would enable predictive applications, such as suggesting the optimal set of characterization techniques for a given material-parameter-quality space, tailored to the user’s research purpose. 
Such databases would make this material knowledge not only visual and interpretable, but also computationally exploitable for planning, automation and AI-assisted workflows.

\subsection{Specific knowledge sublayers}

\begin{figure*}[htpb]
	\centering
	\includegraphics[width=0.95\textwidth]{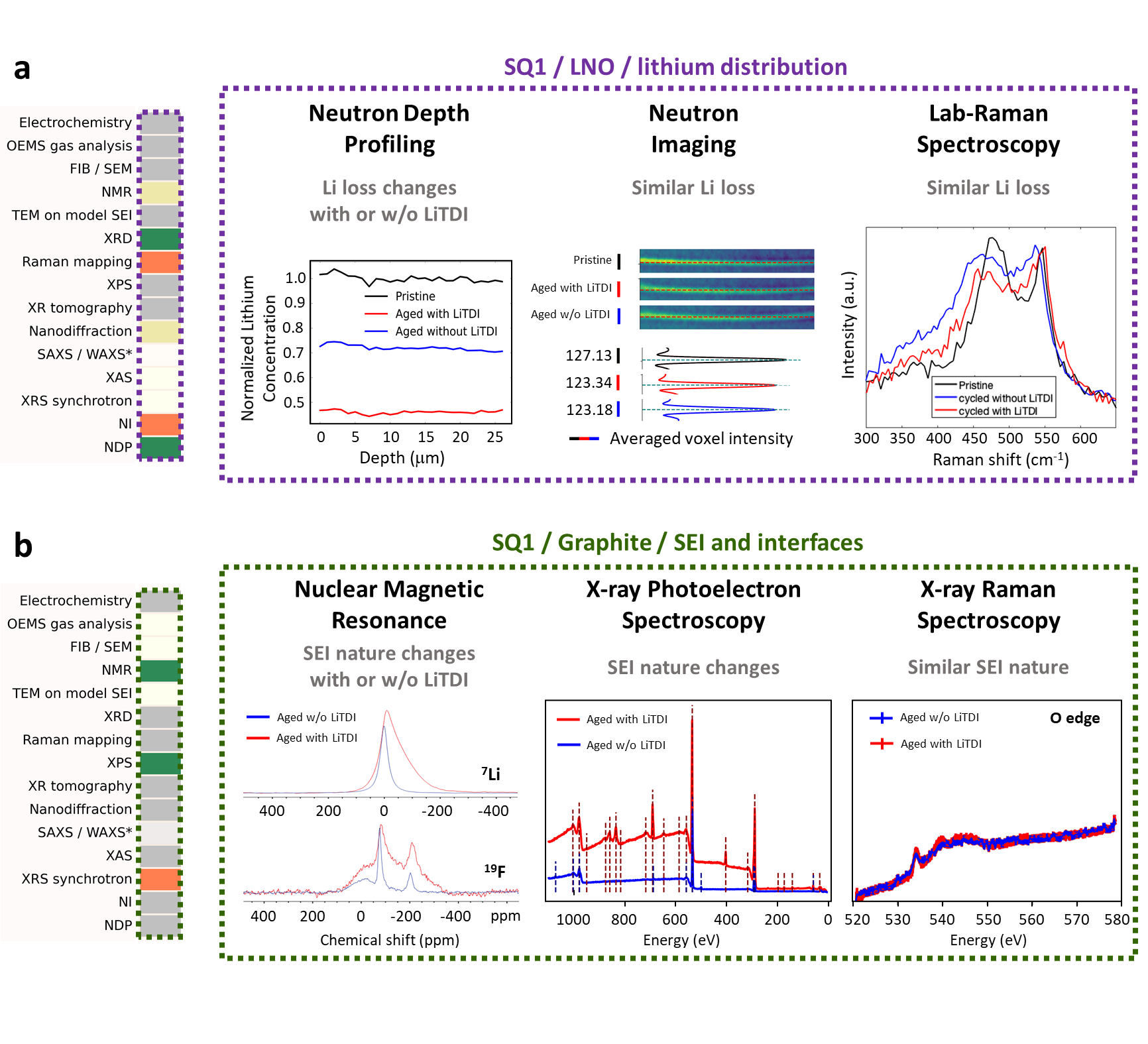}
	\caption{Correlative analysis in subareas of the metaviews. a) Correlative sublayer: SQ1, LNO, Li distribution, probed by Neutron Depth Profiling (NDP, left panel), Neutron Imaging (NI, middle panel), and lab-scale Raman mapping (right panel). %The electrolyte impact on the Li distribution is assessed with varying degrees of sensitivity to changes. 
	In line ``Raman mapping" and column ``Li distribution”, the pixel is orange, as for line ``NI", meaning that no noticeable differences in Li distribution have been obtained using these techniques when comparing LNO electrodes aged with both electrolytes, as seen from the similar attenuated neutron intensities (in red and blue with and without LiTDI respectively) and similar Raman peaks positions and shapes (red and blue spectra). In contrast, NDP revealed changes in the normalized lithium concentration measured in the depth (red and blue data), hence the pixel is colored green. b) Correlative sublayer: SQ1, Graphite, SEI and interfaces investigated with Nuclear Magnetic Resonance (NMR, left panel), lab-scale X-ray Photoelectron Spectroscopy (XPS, middle panel) and synchrotron X-ray Raman Spectroscopy (XRS, right panel). The electrolyte impact on the SEI nature is probed with different sensitivities. In lines ``NMR" and ``XPS", the pixel is green as these techniques evidenced differences in the SEI structure and composition in between graphite electrodes cycled with both electrolytes, seen from changes in the $^{7}$Li and $^{19}$F spectra as well as large-energy band XPS survey spectra. The cell for the ``XRS synchrotron" line is orange as no variations were detected with enough intensity with regards to the measurement error margins, as seen from the O edge data.}
	\label{fig_science_summary_corell}
\end{figure*}

Subareas of the metaview matrix can be isolated to highlight the answers focusing on a given parameterized aspect, as showcased in Figure \ref{fig_science_summary_all}b-c in two areas (dashed purple and green rectangles) and \textbf{Figure \ref{fig_science_summary_corell}} to discuss result correlations. 
The outputs of the corresponding individual datasets can be confronted to learn more on a specific aspect of the chosen material. 
The richness of the scientific interpretations depend on several factors, among them the dataset completeness and depth of analysis. 
Here, as we focus on a proof-of-concept workflow demonstration, not all necessary data were acquired nor analyzed to fully investigate specific aspects. 
In the following, we expand two subareas to illustrate the potentialities of knowledge sublayers analysis, focusing on exemplifying characterization limits and origins of interpretation uncertainty. %\\~\\

\threesubsection{``LNO",``SQ1" (electrolyte impact) and ``Li distribution"}  
The techniques employed were NMR, XRD, Raman mapping, STXM, NI and NDP all applied in ex situ mode (column 3, respectively rows 4, 6, 7, 10, 14 and 15 in Figure \ref{fig_science_summary_all}b, purple dashed rectangle). 
In section \ref{sec_indivDatsets} we have already briefly introduced the results gathered through these techniques, more details being available in Notes 8, 10, 11, 14, 18 and 19, Supporting Information, for each technique respectively. 
Overall there seem to be some level of contradictory answers as mentioned in section \ref{sec_propMapppings} with a collection of ``yes", ``no" and ``unsure" answers (Figure \ref{fig_science_summary_corell}).

Unsure answers were found by NMR and STXM. $^{7}$Li NMR spectrum of the two aged LNO samples shows that the two peaks around 800 and 550 ppm are different, the peak at 550 ppm being more prominent for the electrode aged without LiTDI (\textit{cf.} Figure S10a, Supporting Information). 
The peak at 800 ppm could be attributed to the pristine (fully lithiated) LNO structure while the peak at 550 ppm could be attributed to the LNO monoclinic phase (partially lithiated) (\cite{genreith_probing_2024, li_new_2021, nguyen_new_2024}).
Variations in these two peaks may indicate a partial lithiation of the aged samples with some areas or particles having a lower Li content. However, further expert analysis is needed to ascertain this hypothesis (as notified by the ``unsure" pixel from the expert partner inputs).

Synchrotron nanodiffraction experts also reported uncertain results by STXM on this material and topic, but for a completely different cause, as mentioned before. While the overall in situ experiment monitoring (through combined cell electrochemistry and averaged microdiffraction) indicated expected reaction evolutions (\textit{cf.} Figure S22, Supporting Information, consistent with the electrode-averaged lab XRD results), the analysis of the nanodiffraction data revealed a severe beam damage effect on the locally probed area preventing the electrochemical reaction to occur. It was impossible to analyze the particle-level data, creating the uncertain answer. 

In contrast to NMR and STXM, a positive assessment was achieved with XRD. Experiments on the aged LNO samples showed that, regarding Li distribution, the phase composition and cell parameter dimensions were slightly different (\textit{cf.} Figure S14 and S15, Supporting Information) indicating variations in the Li content distribution in between the electrodes cycled with the two electrolytes. This result was qualitatively corroborated by Neutron Depth Profiling. Figure \ref{fig_science_summary_corell}a (left graph) shows the normalized Li concentration \textit{vs.} the electrode depth measured by Neutron Depth Profiling. 
The overall lithium content is lower for the electrode aged with LiTDI (red line, $\sim$0.46) with respect to the one aged without LiTDI (blue line, $\sim$0.72) with a 36~\% relative decrease.

However, Neutron Imaging and 2D Raman data did not confirm this result (Figure \ref{fig_science_summary_corell}a, middle graph and right panel graph, respectively). 
NI enables the quantification of Li content in ex situ samples from the measured attenuation, yielding similar values on aged LNO independently of the use of LiTDI  ($\sim$123, relative variation of 0.3~\%). Looking at the Raman shifts measured by Raman mapping (Figure \ref{fig_science_summary_corell}a, right graph), the spectra are similar to what is expected from literature on LNO \cite{jacquet_fundamental_2024, flores_in-situ_2018}.
It is noticed that the E$_{g}$ and A$_{1g}$ bands exhibit slight inhomogeneities in band positions, widths and intensity ratios linked to a disordered nature of LNO in its fully lithiated state and different Ni oxidation state ratios. 
These variations are however very similar in between the samples aged with and without LiTDI both intra- and inter- particles, at least at the probed electrode surface, hence the negative answer in the corresponding pixel of the subarea. 
To conclude, both Raman and NI indicate a similar Li loss with or without additive, in contrast to NPD and XRD.\\

By the technique nature, Raman spectroscopy has only access to the material information at the electrode surface illuminated by the monochromatic light beam with a depth on the order of a few 100~nm \cite{jacquet_fundamental_2024} but can capture heterogeneities across this surface providing the probe size (beam spot) is compatible with sample and particle sizes.
NDP is quantitatively sensitive to the Li concentration and has a good depth resolution with a probing length of 26~$\upmu$m in the LNO samples, but averages this depth-resolved information over the surface illuminated by the neutron beam \cite{verhallen_operando_2018, lyons_considerations_2022}.
The NI performed for this workflow has a spatial resolution of 4.2~$\upmu$m in three dimensions over the full electrode for the coin cells but its sensitivity to elements is not monotonic and can greatly vary even between isotopes. It is however more sensitive to the presence of light elements like Li than bulk metallic elements like Ni \cite{lubke_origins_2024, ziesche_neutron_2022, ziesche_multi_2023}, allowing the visualization of Li concentration at least comparatively.
Given the way image analysis was performed for the NI data, only an averaged information over the whole electrode volume was obtained. The full electrode thickness given by the provider is 43~$\upmu$m. 
Hence, NDP is probing more than half its full depth. 
As NI resolution allows an average of 10 voxels in the thickness, NDP has more depth resolution in this situation.
These two techniques can be considered more bulk sensitive than the surface information accessed by Raman spectroscopy.

Therefore, differences in the multi-property metaview column can be explained by i) the difference in probed area in the experiments (bulk \textit{vs.} surface or averaged \textit{vs.} localized), ii) Spatial resolution and iii) sensitivity to the local lithium concentration. 
Material-wise, this ensemble dataset shows that regarding lithium distribution, in electrodes aged with and without LiTDI, the surface facing the separator exhibit a similar nature but that, in depth, cycling with LiTDI induces a higher loss in lithium inventory.

Regarding workflow and technique considerations, we can conclude that, when analyzing data and aiming towards highly correlated ensembles, there are several aspects to consider: i) keep the probed area as a center pivot when comparing results coming from different techniques and ii) take care in assessing the probe interaction with the sample, especially when dealing with highly sensitive and high stakes~/ cost experiment such as the ones performed at synchrotron or neutron facilities. %\\~\\

\threesubsection{``Graphite", ``SQ1" (electrolyte impact) and ``SEI and interfaces"} 
Techniques that yielded results for this combination  were solid state NMR, XPS and synchrotron XRS all performed ex situ (Figure \ref{fig_science_summary_all}c, middle panel, column 7, respectively rows 4, 8 and 13). 
Results obtained from these datasets taken individually were already briefly described in section \ref{sec_indivDatsets} and can be found in Note 8, 12 and 17, Supporting Information. %\ref{Ssec-nmr}, \ref{Ssec-xps} and \ref{Ssec-xrs}

As a reminder, the NMR results pointed towards a more diverse and disordered SEI potentially closer to the graphite surface for the electrode cycled with LiTDI (\textit{cf.} Figure \ref{fig_science_summary_corell}b, left graph). 
Indeed, the $^{7}$Li spectra show a peak under 10~ppm and no peak at around 40~ppm, which would indicate a mostly delithiated structure \cite{andersen_strategies_2021, gotoh_mechanisms_2020, markert_lithium_2025}.
The peak under 10~ppm is around 0~ppm for the electrode aged without LiTDI, while it is shifted towards more negative values with a broader tail for the electrode aged with LiTDI.
This is the main indication towards the greater SEI environment variety and its more disordered structure.
Regarding the $^{19}$F spectra, there are two noticeable peaks at -75~ppm and -205~ppm, respectively corresponding to LiPF$^{-}_{6}$ and LiF \cite{haber_what_2018}.
The peaks for the sample aged with LiTDI are, as for $^{7}$Li, broader, indicating either more disordered environments or closeness to the graphitic surface.

Similarly, XPS spectra (\textit{cf.} Figure \ref{fig_science_summary_corell}b, middle graph) were different for the two aged electrodes and more complex for the electrode aged with LiTDI with more peaks and more intense ones (Figure S18a, Supporting Information). %\ref{Ssfig_XPS_survey}
Looking at Figure \ref{fig_science_summary_corell}b, all the peaks (but one around 1100~eV) from the electrode aged without LiTDI are shared with the one aged with LiTDI.
This shows that the compounds detected in the SEI formed without LiTDI are also present in the SEI formed in presence of LiTDI.
Moreover, the appearance of more peaks for the sample cycled with LiTDI indicates the presence of more compounds in the LiTDI induced SEI and the higher peak intensity can be imputed to a higher concentration of the related compound cause by the LiTDI additive in the electrolyte.

In contrast, synchrotron XRS could detect some signal in the O K-edge (\textit{cf.} Figure \ref{fig_science_summary_corell}b, right graph) attributed to the Li$_{2}$CO$_{3}$ SEI component \cite{hernandez_understanding_2025} in both electrodes.
However, the variations between the electrodes aged with and without LiTDI were within the experimental resolution (Figure S25b, Supporting Information).\\ 

The NMR and XPS results consistently converge to a chemically more diverse SEI as a result of cycling graphite electrodes with LiTDI.
NMR also suggests that this SEI is thinner, which is consistent with the findings of the TEM study on model LNO samples showing a non-honomgeneous SEI with thinner regions on the Ni particles in presence of LiTDI (\textit{cf.} Figure S12, Supporting Information). %\ref{Sfig_TEM_results_composite}
This variation in the SEI layer thickness could be caused by its more complex composition inducing some areas with a more unstable composition breaking more easily upon cycling. 
The fact that XRS does not capture differences between the aged electrodes is probably due to measurement uncertainties and its bulk nature compared to a surface-sensitive technique like XPS.
Indeed, the XPS only probes depths up to around 10~nm which means that the signal comes mostly, if not uniquely, from the SEI layer, while XRS sums information coming from the graphite particles, binder and other phases. 
Moreover, C and O edges are dominated by particles and binder, rendering SEI species detection difficult.
Measuring at SEI-specific edges (as, for instance, Li or F) can be attempted but necessitates large counting times as signals are very small. 
In this regard, NMR is advantageous as chemical shifts enable to isolate some specific chemical environments even present at very low concentrations, specially when using magic angle spinning detection in ex situ samples.

In conclusion, the subarea analysis shows that the addition of LiTDI in the electrolyte has an important impact on the SEI nature and structure, despite not leading to noticeable improvements on the cycling lifespan of LNO~/ Gr cells. 
In this case, we can hypothesize the interplay of opposite effects, \textit{e.g.} the degraded~/ more complex SEI formed in LiTDi cells is probably counterbalanced by more stable graphitic conductive pathways in graphite and higher Ni oxidation state in LNO.
In this example, again, sensitivity and the probed area are shown to be key parameters to consider when analyzing multi-technique results. 
There are not real contradictions among the technique-dependent results. They rather provide different perspectives on the same system, offering complementary information that would likely have been missed otherwise. 
A subset of information can also guide the interpretation of data from other techniques with comparable spatiotemporal resolutions (for instance, XPS \& FTIR, tomography \& NI \& NDP,...) but different scopes, such as TEM results providing complementary insights into the NMR analysis.

\section{Challenges and perspectives}
	
\begin{figure*}[!htpb]
		\centering
		\includegraphics[width=0.85\linewidth]{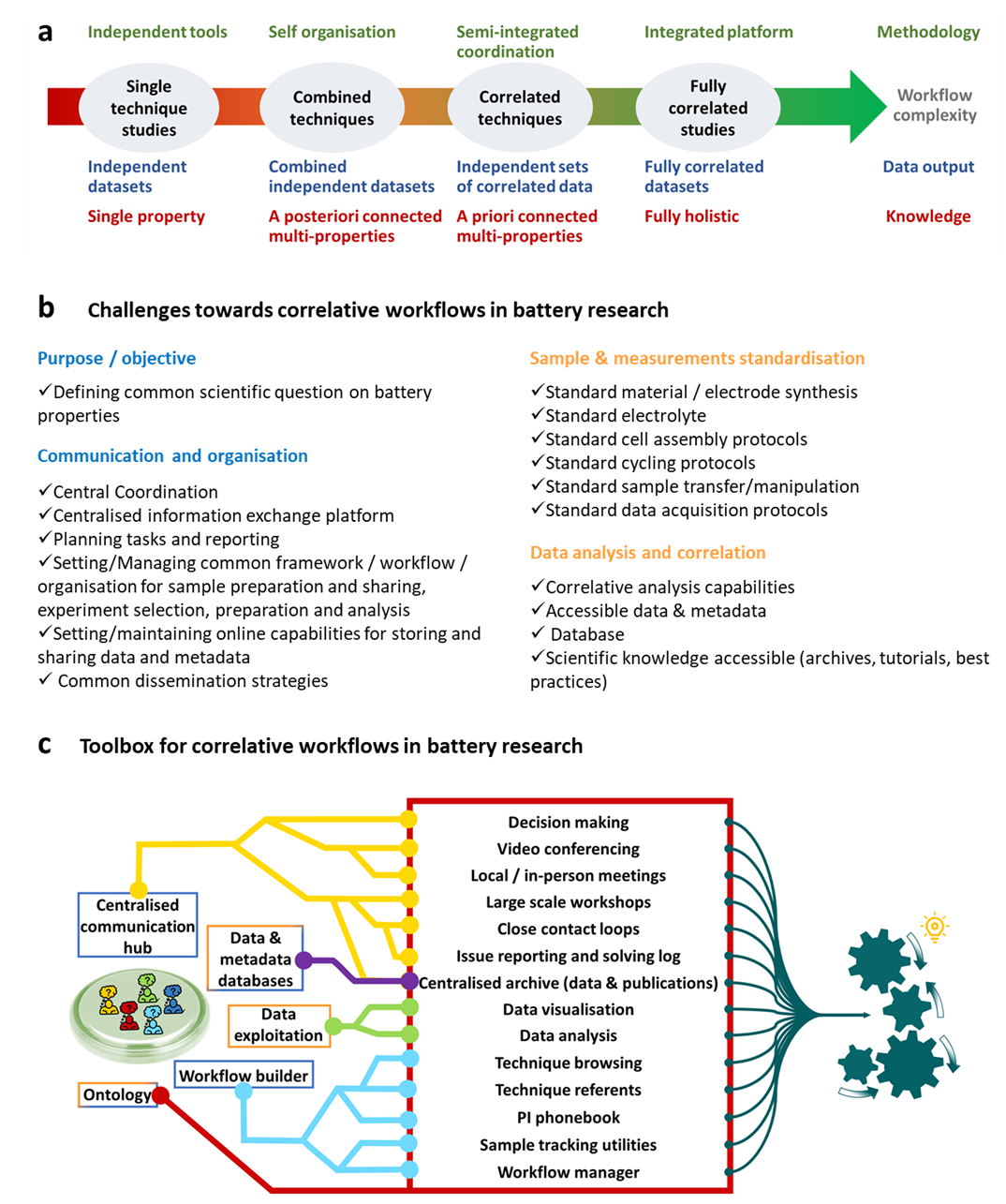}
		\caption{Roadmap towards fully holistic characterization workflows. a) Stages of workflow complexity. Proof-of-concept of Stage III (``Correlated techniques") is reported in this work. Stage IV necessitates to overcome challenges (b) and implement an array of tools (c). Challenges encompass communication, organization, standardization and data correlation aspects. A Battery Characterization Platform toolbox should enable users to conceive and execute workflows, with easy access to input~/ output knowledge and data. Workflow builder, databases, data infrastructures and ontology are coordinated via a centralized communication hub. }
		\label{fig_challenges}
	\end{figure*}

Widespread studies and holistic understanding are out of reach for a single researcher working alone, as well as it is for a single laboratory. 
However, the more people are partnering up to focus on a dedicated scientific subject, the more difficult it is to develop efficient and constructive ways to correlate data and results.
Therefore, complex workflows - as the one demonstrated here - are needed to shift paradigm from stand-alone isolated battery characterizations to more integrated platforms, where holistic investigation is pursued and prioritized (\textit{cf.} \textbf{Figure \ref{fig_challenges}a}).
	
Many challenges are to be faced when deploying a large-scale workflow infrastructure to produce extended data correlations (\textit{cf.} \textbf{Figure \ref{fig_challenges}b}). 
Sharing, communication, and synchronization capabilities are key to define the common purpose and objective, and further execute a successful multi-site measurement campaign. 
In this work, the organization was managed and centralized ``manually" by a few people, thanks to the solid framework provided by such a large EU-funded project as BIG-MAP in Battery 2030+.
Such project enabled the conception and development of standardized sample preparation and measurements protocols, as well as common tools for ontologized metadata production, data storing on common servers, data access protocols and data analysis descriptions, including tutorials and best practices, made available to all partners. 
	
The proof of concept workflow reported here allowed to identify the many challenges that need to be tackled now to expand the concept and create more autonomous correlative workflows and a generalized European battery characterization platform, relying on a toolbox as summarized in \textbf{Figure \ref{fig_challenges}c}.
	
First of all, a communication hub must be set between the workflow partners, ensuring fluid communication. 
A centralized and normalized protocol should be established to ease contact among partnering facilities (either labs or Large Scale Facilities). 
This hub should be a platform where users or researchers can send requests for information or execution to each partner (who agrees to provide access (or discuss access modalities) to their suite of experimental techniques). 
Formalized contact form, general information, standard protocols, chat~/ video conferencing capabilities, read-only accessibility planning, \textit{etc.} are among the necessary tools for effective exchanges and action programming. 
The communication hub should also direct users and researchers to the various platform tools available on-line or via Apps. 
	
It is critical to know what instruments are available for battery characterizations in the platform, with important details such as experimental setup characteristics and constraints, sample environments, as well as practical information (where are the instruments located, when can they be used, who to reach out to plan and~/ or perform experiments, what are the safety rules to comply with, \textit{etc.}). 
A visualization and browsing tool must be available to ease technique visibility and accessibility. 
Moreover, a workflow builder tool must be developed to enable users or researchers to access the capabilities of the platform and build an effective, optimized and comprehensive workflow suited to their request, \textit{e.g.}, proposing a coordinated sequence of characterizations capable to answer their scientific questions in a specific R\&D context. 
This workflow builder is to be connected to the technique visualization \& browsing tool, as well as to data \& metadata databases, data exploitation infrastructure and ontology.
	
Databases are input and output services necessitating continuous refinement and expansion throughout the whole platform operating lifespan. 
Uploading and downloading datasets generated by the workflow should ideally be enabled by a centralized sharing tool, with automatic access to all partners.
A centralized European archive would be also necessary to include access to previously generated open access datasets (and attached papers), relevant to any inquiry.
The BIG-MAP archive is a good starting point focused on dataset storage and sharing capabilities.
	
Data exploitation infrastructure relies, first, on built-in sample tracking utilities, as sample ID, metadata, experiment DOIs, linked experiment libraries, shipment details, \textit{etc.}.
All these information might be needed within a running workflow but also later for reproducibility check or for reusing results in subsequent workflows, possibly beyond the original team~/ project. 
To facilitate specific exploitations, the common platform should also include links and contacts to external analysis software, alternative protocols, expert analysts and theory partners, thereby ensuring a continuous enhancement of pertinent and efficient data analysis capabilities.
	
Furthermore, every aspect of the workflow value chain must be ontologized.
Whether from services provided to education and dissemination or from generated datasets to scientific interpretation, all of this is key to make the workflow building and platform usage remotely possible.
Ontology is not only mandatory to ensure internal cooperation and integration within an existing framework, but it is required to ease the insertion of new partners and experimental techniques to expand the influence, usage and applicability of the European Experimental Platform.
The development of tools like the BattInfo ontology \cite{clark_toward_2022} and the Onterface platform 
\cite{stier_onterface_nodate} (ontologized interface for process automation) need to be continued and expanded to all aspects of data acquisition and correlation. 
This will require a continuity of efforts in the future to develop, deliver and maintain ontologized tools targeted to generic usage (with needs to allocate and manage manpower, budget and infrastructures).

More infrastructures, tools and enablers are needed to facilitate large scale collaborations aiming at establishing and using large dataset ensembles, targeting the final step of a correlated holistic study (\textit{cf.} Figure \ref{fig_challenges}c).
These modalities are yet to be developed and should be integrated in a full fledged workflow, as they are intricately intertwined. 
Building databases, libraries and statistics from different sources are needed in the future to accelerate battery material and technology discoveries.
Regarding experimental characterizations, fast screening and in-depth analysis are two faces of the same medal.
It is necessary to identify key techniques or general observables allowing for a fast and simple first screening, thereby enabling rapid discrimination before setting more detailed investigations with focused objectives. 
The latter are possibly dictated by the outputs from the first screening via intelligent feed-back loops, potentially autonomous using AI power.
An early demonstration of a closed-loop, structure-targetted, autonomous experimentation was recently performed for nanoparticle synthesis \cite{anker_autonomous_2025} that is relevant to stage IV.
However, fully hollistic approaches are envisioned theoretically but not yet deployed in practice, a key challenge being the execution of thorough data correlation during the analysis.
	
Multi-property metaviews (Figure \ref{fig_science_summary_all}b-c) are suited patterns to merge interpretations and provide quick visual evidence of results heterogeneities due to spatially~/ temporally~/ methodologically biased findings. 
In this early stage development, they are not sufficient to gain complete and in-depth detailed and correlated insights into the material behavior and related degradation mechanisms. 
Yet, they constitute a first semi-integrated action towards fully autonomous platforms.

To summarize along with Figure \ref{fig_challenges}, efficiently leveraging coordinated characterization workflows is facing numerous challenges.
Some are related to the practical and technical implementation and execution of such endeavors.
However, these aspects appear non-essential if the huge amount of data generated cannot be properly correlated and analyzed. 
Unfortunately, no tools seem to be existing so far to efficiently leverage this critical aspect. 
These two categories of challenges appear to be of paramount importance to establish a functioning and efficient pan-European Experimental Platform for battery characterization.
	
\section{Conclusion}

	In this paper, we demonstrated the capability to conduct timely coordinated experiments involving the collaboration of 15 partners gathering research and technology organizations, academic laboratories, and several synchrotron and neutron Facilities. 
	This large scale multimodal workflow enabled characterizing the properties of two selected Li-ion battery chemistries (Gr~/ LNO electrodes, electrolytes with or without LiTDI additive) cycled in a standardized full cell configuration, using multi-site, multi-technique and multi-scale coordinated experiments.
	The importance of a meticulous definition of suitable scientific question(s) as well as protocols for sample standardization serves as the backbone to a vast characterization campaign integrating reliability and reproducibility. 
	The use of common tools such as the BIG-MAP Notebook and the BIG-MAP Archive played a crucial role in facilitating the effective storage and sharing of data and metadata, a key requirement to enable efficient integration and analysis. 
	Moreover, efficient collaborative frameworks using standardized language (via ontologized and shared tools) was instrumental to effectively validate results across participants and users. 
	
	Through a combination of different probes, sensitivity and resolutions (in both time and space), such large scale workflows can allow for holistic views into the system mechanisms and nature by leveraging accurate data correlation. 
	We introduced collaborative tools in the form of visual patterns, or two-dimensional technique-property genotypes, that represent specific regions of the explored sample-question-characterization space.
	However, while data acquisition steps and workflow organization has been successfully validated, fully correlative data analysis remains a critical challenge, as robust self-regulated methods to handle multiple large and heterogeneous datasets are lacking. 
	The workflow has paved the way for the European Battery research community to engage in synergistic collaborative research and development and has initiated the implementation of a European multi-modal experimental platform for which the next challenges include leveraging automation of multimodal data cross-analysis and autonomous interpretation.\\

\clearpage

\medskip
\textbf{Supporting Information} \par %Please delete the Suppporting Information statement if it is not applicable. Please supply Supporting Information in another file. Supporting information should not be provided in .tex format
Supporting Information is available from the author.

% Acknowledgements
\medskip
\textbf{Acknowledgements} \par %delete if not applicable))
This project has received funding from the European Union’s Horizon 2020 research and innovation programme under grant agreement No 957189.
The project is part of BATTERY 2030+, the large-scale European research initiative for inventing the sustainable batteries of the future.
Beamtime at the ESRF was granted within the Battery Pilot Hub MA4929 “Multi-scale multi-techniques towards an European battery hub”.
The authors thank BIG-MAP partners at WUT for providing the LiTDI additive.
T.V. also acknowledges The Pioneer Center for Accelerating P2X Materials Discovery (CAPeX),  DNRF Grant number P3, for hosting the BIG-MAP archive.

% Conflict of interest
\medskip
\textbf{Conflict of interest} \par
There is no conflict of interest to declare.

% Author contributions
\medskip
\textbf{Author contributions} \par
F. C. and S. L. conceptualized, built and coordinated the workflow as well as the data analysis at the correlative and metaview levels. C. Herrera participated to the building and coordination of the workflow. All authors contributed to the experimental acquisition and the individual dataset analysis for their respective techniques. F.C. and S.L. wrote the manuscript, further reviewed by all co-authors.

% Data availability
\medskip
\textbf{Data availability statement} \par
Data used and acquired in this work is available upon request to the authors. Data acquired at the ESRF are available under proposal MA4929.

% References
\medskip

% Use the following code if you wish to generate your bibliography with BibTeX;
% replace the string "MSP-template" below with the name(s) of
% the BibTeX data base(s) you want to use.
% The resulting bibliography-output (the content of the .bbl file)
% must be pasted back into this file before submission.
% Please also include your BibTeX data base file(s) in your submission
% so that we can re-run BibTeX if necessary.
%
%\bibliographystyle{MSP}
%\bibliography{WorkflowMod.bib}

\clearpage
% Table of contents entry should be 50 - 60 words long
% Image should be 55 mm broad and 50 mm high or 110 mm broad and 20 mm high

\begin{figure}
\textbf{Table of Contents}\\
\medskip
  \includegraphics[width=55mm, height=50mm]{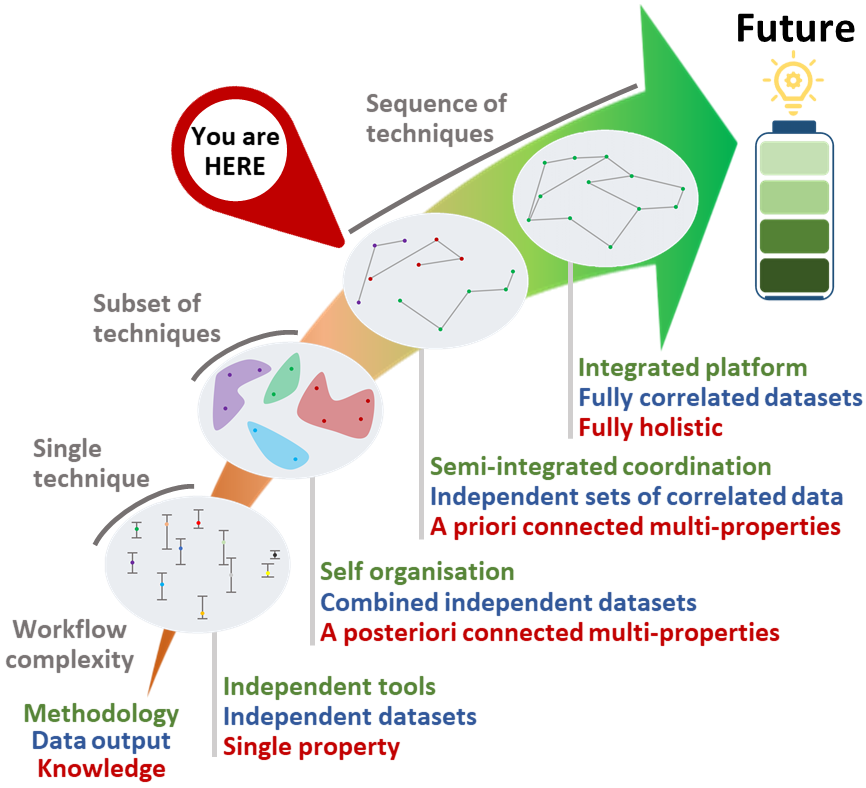}
  \medskip
  \caption*{Roadmap for reaching integrated and holistic large-scale multi-modal battery characterization workflows. The concept and practical implementation of a semi-integrated coordinated workflow is reported here. It allows interpretations from a stand-alone technique level to multi-property overviews and correlative analysis on specific subsets. An analysis is made of the challenges to overcome for evolving towards a fully holistic integrated platform.
  }
\end{figure}

\end{document}